%% file: Bohrv2.tex
\documentclass[11pt]{article}
\usepackage{amssymb}
\usepackage{amsmath}\usepackage{graphicx}  

\input{Bohrificationpreamble.tex}

\newcommand{\mmp}{measurement problem}
\newtheorem{Definition}{Definition}[section]

\newtheorem{Theorem}[Definition]{Theorem}

\newtheorem{Corollary}[Definition]{Corollary}
\topmargin = - 1 cm \textheight = 23 cm \textwidth = 15 cm
\begin{document} 
\pagenumbering{arabic} \setlength{\unitlength}{1cm}\cleardoublepage
\thispagestyle{empty}
\title{\Large{Bohrification: From classical concepts to commutative algebras}}
\author{Klaas Landsman\footnote{Institute for Mathematics, Astrophysics, and Particle Physics, Faculty of Science, Heyendaalseweg 135, 6525 {\sc aj Nijmegen, The Netherlands}, 
Radboud University Nijmegen,
\texttt{landsman@math.ru.nl}}}
\date{\today}\vspace{-5mm}
\maketitle\vspace{-10mm}
\begin{center}
\emph{Dedicated to the memory of Rudolf Haag (1922--2016)}
\end{center}
 \begin{abstract} 
\noindent 
The Bohrification program is an attempt to interpret Bohr's mature doctrine of classical concepts  as well as his earlier correspondence principle in the operator-algebraic formulation of quantum theory pioneered by von Neumann.
 In particular, this involves the study of commutative \ca s in relationship to noncommutative ones. This relationship 
 may  take the form of either \emph{exact} Bohrification, in which one studies commutative unital C*-subalgebras of a given noncommutative C*-algebra, or  \emph{asymptotic} Bohrification, which involves deformations of commutative \ca s into noncommutative ones. Implementing the doctrine of classical concepts, exact Bohrification  is an appropriate framework for the Kochen--Specker Theorem and for (intuitionistic) quantum logic, culminating in the topos-theoretic approach to \qm\ that has been developed since 1998. 
Asymptotic Bohrification was inspired by the correspondence principle and forms the right conceptual and mathematical framework for the explanation of the emergence of the classical world from quantum theory (incorporating the
 measurement problem and the closely related issue of spontaneous symmetry breaking). 
 The  Born rule may be derived  from both exact and  asymptotic Bohrification, which  reflects
 the Janus faces of probability as applying to individual random events and to relative frequencies, respectively.

We review the history, the goals, and the achievements of this program so far. 
\end{abstract}
\maketitle
\begin{small}
\begin{quote}
 \tableofcontents
\end{quote}
\end{small}
\clearpage
\newpage
\section{Introduction}\label{sec1}
Despite considerable research (e.g.\ Joos et al, 2003; Schlosshauer, 2007; Landsman, 2007; Wallace, 2012), it remains  unclear how
quantum mechanics---accepted as the fundamental theory of the physical world---gives rise to  classical physics, or at least to the appearance thereof. The relationship between classical and quantum physics played a fundamental role in the work of Niels Bohr, who contributed two important ideas (or perhaps ideologies):
\begin{itemize}
\item The  \emph{correspondence principle} (Bohr, 1976),  which---originally in the context of atomic radiation---suggested an asymptotic relationship between classical and quantum mechanics (see also Mehra \& Rechenberg, 1982;  Hendry, 1984; Darrigol, 1992);
\item The \emph{doctrine of classical concepts}, which Bohr succinctly summarized  as follows:
 \begin{quotation}
 `However far the phenomena transcend the scope of classical physical explanation, the account of all evidence must be expressed in classical terms. (\ldots) The argument is simply that by the word \emph{experiment} we refer to a situation where we can tell others what we have done and what we have learned and that, therefore, the account of the experimental arrangements and of the results of the observations must be expressed in unambiguous language with suitable application of the terminology of classical physics.' (Bohr, 1949, p.\ 209). 
\end{quotation}
In brief, the only way to look at the quantum world is through classical glasses. 
See also Bohr (1985, 1996), as well as Scheibe (1973), Folse (1985),  Camilleri \& Schlosshauer (2015), and Zinkernagel (2016) for further analysis  (which we recommend but shall not review, Bohr's role in this work being that of an inspirator rather than a subject of exegesis---we will similarly not discuss (anti-)realism etc.). 
\end{itemize}
The tension between these ideas was  well  captured by Landau \& Lifshitz (1977, p.\ 3):
 \begin{quotation}
`Thus \qm\ occupies a very unusual place among physical theories: it contains classical mechanics as a limiting case, yet at the same time it requires this limiting case for its own formulation.' 
\end{quotation} 
What adds to the difficulties is the need to combine conceptual and mathematical analysis, but precisely at this point the approach we advocate promises a clean separation between the two:  the necessary conceptual analysis has largely been performed by Bohr,  whilst von Neumann's theory of operator algebras---as extended to \ca s (Gelfand \& Naimark, 1943) and subsequently to
noncommutative geometry (Connes, 1994)---provides the required mathematical tools. What has been lacking so far is a relationship between the two, and this is what our ``Bohrification'' program intends to provide. Indeed, we feel that 
 each author  on his own failed to solve the foundational problems of \qm:  Bohr's  conceptual ideas must first be interpreted within the formalism of \qm\ before they can be applied to the physical world (an intermediate step Bohr himself seems to have considered of little interest, if not wholly superfluous), whereas conversely, von Neumann's brilliant mathematical work needs to be supplemented by sound conceptual moves before it, too, may be related to actual physics. Von Neumann (1932) did write down the most general form of the Born rule, which is an important link between physics and the  \Hs\  formalism of \qm\ he developed, but this is not nearly enough;  what we propose is that Bohr provided these conceptual moves.
 
In this respect, Bohr and von Neumann historically played asymmetric roles: whereas Bohr hardly attempted to to relate his ideas to the mathematical formalism of \qm\ (not even to Dirac's),  von Neumann (1932) famously did spend a great deal of attention to conceptual issues in \qm. Although this work  was extremely influential, and von Neumann's  analysis of  hidden variables did set the stage for part of our Bohrification program (albeit in a contrapositive way), his work on entropy or the measurement problem has had little influence on the program proposed here.  
 In any case,  von Neumann never advocated the particular use of commutative  operator algebras we now see as the basis of the relationship between his mathematics and the real world---seen through the classical glasses suggested  by Bohr's doctrine. 
 In any case, although  Bohr and von Neumann tend to be jointly bracketed  under the Copenhagen Interpretation---a notion to be treated with due reservation (Faye, 2002; Howard, 2004; Camilleri, 2007)---a synthesis of their ideas on \qm\ is difficult to find in earlier literature.
The central idea of Bohrification is to interpret Bohr's  doctrine mathematically, as follows: 
\begin{trivlist}
\item[] \emph{In so far as their physical relevance is concerned, the noncommutative operator algebras of quantum-mechanical observables are accessible only through commutative algebras.}
\end{trivlist}

 On this interpretation, the Bohrification program splits into two parts, distinguished by the specific relationship between a given 
 noncommutative operator algebra $A$ (whose self-adjoint elements mathematically represent the observables of some quantum system) and the commutative algebras that supposedly give access to $A$. Either way, we assume that $A$ is a \ca: this is an associative algebra (over $\C)$ equipped with an involution (i.e.,  a real-linear map
$A\rightarrow A^*$ such that  $a^{**} =  a$, $(ab)^*  = b^* a^*$, and $(\lm a)^*  =
\ovl{\lm}a^*$ for all $a,b\in{A}$ and $\lm\in\C$), as well as a norm in which $A$ is complete (i.e., a Banach space),  such that algebra, involution, and norm are related by the axioms $\|ab\|\,\leq\, \| a\|\,\| b\|$ and $\|a^*a\|  =  \| a\|^2$. The two main classes of \ca s are:
\begin{itemize}
\item The space $C_0(X)$ of all  continuous functions $f:X\raw\C$  that vanish at
infinity (i.e., 
 for any  $\varep>0$ the set $\{x\in X\mid |f(x)|\geq \varep\}$ is compact),  where $X$ is some locally compact Hausdorff space, with pointwise  addition and
multiplication, involution  $f^*(x)=\ovl{f(x)}$,
and a  norm $\|f\|_{\infty}=\sup_{x\in X} \{|f(x)|\}$.
It is of fundamental importance for physics and mathematics that
 $C_0(X)$ is \emph{commutative}. Conversely,  Gelfand \& Naimark (1943) proved that every commutative \ca\ is isomorphic to $C_0(X)$ for some  locally compact Hausdorff  space $X$, which is uniquely determined by $A$ up to homeomorphism ($X$ is called the \emph{Gelfand spectrum} of $A$). 
 Note that $C_0(X)$ has a unit (namely the function $1_X$ that is equal to 1 for any $x$) iff $X$ is compact.
   \item 
   Norm-closed subalgebras $A$ of the space $B(H)$  of all bounded operators on some \Hs\ $H$
   for which $a^*\in A$ iff $a\in A$; this includes the case $A=B(H)$. Here one uses  the standard operator norm $\|a\|=\sup\{\|a\psi\|, \ps\in H, \|\psi\|=1\}$,  the algebraic operations are the natural ones, and the
 involution is the adjoint (i.e., hermitian conjugate). 
  If $\dim(H)>1$, $B(H)$ is a  \emph{non-commutative} \ca. 
  
  An important special case is the \ca\ 
 $B_0(H)$ of all \emph{compact} operators on  $H$, which fails to have a unit whenever $H$ is infinite-dimensional 
 (whereas $B(H)$ is always unital). In their fundamental paper, Gelfand \& Naimark (1943) also proved that every  \ca\ is isomorphic to $A\subset B(H)$ for some  \Hs\ space $X$.
\end{itemize}
These two classes are related as follows: in the commutative case $A= C_0(X)$, one may 
 take $H=L^2(X,\mu)$ (where the measure $\mu$ is supported by all 
 of $X$), on which $C_0(X)$ acts by multiplication operators, that is, $m_f\psi=f\psi$, where  $f\in C_0(X)$ and $\psi\in L^2(X,\mu)$.

\ca s were introduced by  Gelfand \& Naimark (1943), generalizing the rings of operators studied by von Neumann  during the period 1929--1949, partly in collaboration with Murray 
 (von Neumann, 1929, 1931, 1938, 1940, 1949; Murray \& von Neumann, 1936, 1937, 1943). These rings 
 are now aptly called \emph{\vna s}, and arise as the special case where a \ca\ $A\subset B(H)$ satisfies $A=A''$, in which for any subset $S\subset B(H)$ we define the \emph{commutant} of $S$ by $S'=\{a\in B(H)\mid ab=ba\,\forall\, b\in S\}$, and
the \emph{bicommutant} of $S$ as  $S''=(S')'$. A \emph{state} on a \ca\ $A$  is a linear map $\om:A\raw\C$ that is \emph{positive} in that $\om(a^*a)\geq 0$ for each $a\in A$, and \emph{normalized} in that, first noting that positivity implies boundedness, $\|\om\|=1$, where $\|\cdot\|$ is the usual norm on the Banach dual $A^*$ (if $A$ has a unit $1_A$, then the latter condition is equivalent to
$\om(1_A)=1$). The Riesz--Radon representation theorem in measure theory gives a bijective correspondence between states
 $\om$ on $A=C_0(X)$ and  probability measures $\mu$ on $X$,  given by $\om(f)=\int_X d\mu\, f$, for any  $f\in C(X)$.
If $A=B(H)$, then any density operator $\rh$ on $H$ gives a state $\om$ on $B(H)$ by $\om(a)=\Tr(\rh a)$, but if $H$ is infinite-dimensional there are interesting other states, too, cf.\ \S\S\ref{GKS},\ref{KSC}.
 See Kadison \& Ringrose (1983, 1986) for a first introduction to operator algebras.

The use of \ca\ and \vna s in mathematical physics has been widespread for over half a century now,  typically in the context of quantum systems with infinitely many degrees of freedom (Bratteli \&  Robinson, 1981; Haag, 1992). More recently, it has become clear that they also form an effective tool when other limiting operations play a role, such as $\hbar\raw 0$, where $\hbar$ is Planck's constant (Rieffel, 1989, 1994; Landsman, 1998), see also below.  Returning to our main topic of Bohrification, both of these applications of \ca s have now found a common conceptual umbrella:  
 \begin{itemize}   
 \item  \emph{Asymptotic  Bohrification} looks at commutative C*-algebras asymptotically included in $A$ by a deformation procedure involving so-called continuous fields of C*-algebras (see appendix). As we shall see, this covers the Born rule (in its frequency interpretation), 
  the measurement problem (which we  relate to  the  classical limit of \qm), and the closely related issue of spontaneous symmetry breaking. All of this may be bracketed under the general theme of emergence
  (Landsman, 2013).
\item  \emph{Exact  Bohrification}  studies commutative C*-algebras contained in $A$, assembled into a poset (i.e., partially ordered set) $\CA$ whose partial ordering is given by set-theoretic inclusion (for technical reasons we assume that $A$ has a unit $1_A$ and that each commutative C*-subalgebra $C\in\CA$ of $A$ contains $1_A$). Applications to the foundations of \qm\ so far include the  Born rule (for single experiments), the Kochen--Specker Theorem, and (intuitionistic) quantum logic.  The topos-theoretic approach to \qm\ (which incorporates the last two themes just mentioned) is heavily based on $\CA$, and also, in a purely mathematical setting, $\CA$ turns out to be a useful new tool for the classification of \ca s.
\end{itemize}

 The conceptual roots of the exact Bohrification program  lie in Bohr's doctrine of classical concepts, but the mathematical interpretation of this doctrine as proposed here starts with
 von Neumann's (1932) proof that (in current parlance) non-contextual linear hidden variables compatible with  \qm\ do not exist;
 see Caruana (1995) for historical context.  Von Neumann's  assumption of linearity of the hidden variable on all operators was criticized by Hermann (1935), whose work was unfortunately ignored, and subsequently by  Bell (1966), whose paper  became well known.  Both Bell (1966) and Kochen \& Specker (1967) then obtained a similar result to von Neumann's on the assumption of linearity merely on \emph{commuting} operators. 
 Furthermore,  responding to different ideas in von Neumann (1932), the mathematician Mackey (1957) introduced probability measures $\mu$ on the closed subspaces of a \Hs, based on an additivity assumption on mutually orthogonal subspaces.
 These formed the basis of his colleague Gleason's (1957) deservedly famous characterization of such measures in terms of density operators.
 
 As we shall explain in \S\ref{GKS},  \emph{commutative} C*-subalgebras of $B(H)$ form an appropriate context for Mackey, Gleason, and Kochen--Specker; this shared background is no accident, since the  Kochen--Specker Theorem is a simple corollary of Gleason's. Moreover, Kadison \& Singer (1959) studied pure states on \emph{maximal} commutative C*-subalgebras of $B(H)$, launching what became known as the \emph{Kadison--Singer conjecture}, which we also regard as a founding influence on  the (exact) Bohrification program.  
Two decades later, the reinterpretation of the  Kochen--Specker Theorem in the language of topos theory led Isham \& Butterfield (1998), Hamilton, Isham, \& Butterfield (2000), and D\"{o}ring \& Isham (2010)  to  a poset similar to $\CA$ in the setting of \vna s (though with the opposite ordering). Finally, $\CA$ was introduced by Heunen, Landsman \& Spitters (2009), again in the context of topos theory. It soon became clear that even apart from the latter context $\CA$ is an interesting object on its own, especially as a tool for analyzing C*-algebras. 
 
 Asymptotic Bohrification conceptually originated in Bohr's correspondence principle, which originally related the statistical results of quantum theory to the classical theory of radiation in the limit of large quantum numbers  (Bohr, 1976), but was later generalized to the idea that classical physics should arise from quantum physics in some appropriate limit (Bokulich, 2008).
 Mathematically, the correspondence principle eventually gave rise to the field of \emph{semiclassical analysis} (Ivrii, 1998; Martinez, 2002), and, more importantly for the present paper, to a
  new approach to quantization theory developed in the 1970s under the name of \emph{deformation quantization} (Berezin, 1975; Bayen et al, 1978).  Here  the noncommutative algebras characteristic of \qm\ arise as deformations of so-called Poisson algebras; the underlying mathematical idea ultimately goes back to Dirac (1930). In Rieffel's (1989, 1994) approach to deformation quantization, the algebras in question are \ca s and hence the entire apparatus of operator algebras---and of  the closely related field on noncommutative geometry (Connes, 1994)---becomes available. 
  
  This led to a general operator-algebraic approach to (deformation) quantization and the classical limit of \qm\ (Landsman, 1998), the scope of which was substantially enlarged in Landsman (2007, 2008) so as to deal with both the macroscopic limit  (i.e., $N\raw\infty$, which also incorporates Bohr's original limit of large quantum numbers)
and the semiclassical limit ($\hbar\raw 0$) at one go. This, in turn, led to a new approach to the measurement problem (Landsman \& Reuvers, 2013) and to spontaneous symmetry breaking (Landsman, 2013), which we now recognize as instances of  asymptotic Bohrification.

The plan of this paper is as follows. We first elaborate on the path alluded to above from von Neumann (1932) to the poset $\CA$ and hence to exact Bohrification. Next, 
 the Born rule provides a nice context for a first acquaintance with the Bohrification program in its two variants (i.e., exact and asymptotic), as it falls under both headings and hence may be understood from two rather different points of view.
After an intermezzo on  conceptual issues with limits in the context of  asymptotic  Bohrification, we  end with an overview of our strategy for first (re)formulating and then solving the measurement problem. 

This paper mainly concentrates on the Bohrification program in  physics. For applications to mathematics see  D\"{o}ring \& Harding (2010), Hamhalter (2011), Hamhalter \& Turilova (2013), Wolters (2013),   D\"{o}ring (2014), Heunen (2014a,b),  Lindenhovius, 2016), as well as the forthcoming review by
 Landsman \& Lindenhovius (2016).
  \section{Gleason, Kochen--Specker, and commutative \ca s}\label{GKS}
Let $H$ be a \Hs.  Von Neumann (1932) introduced the  concept of an \emph{Erwartung} (i.e., \emph{expectation value}) as a linear map $\om':B(H)_{\mathrm{sa}}\raw\R$ that satisfies $\om'(a^2)\geq 0$ for each $a\in B(H)_{\mathrm{sa}}=\{a\in B(H)\mid a^*=a\}$, as well as $\om'(1_H)=1$.
Using the canonical decomposition $a=a'+ia''$, where now $a\in B(H)$ and 
  $a'= \half (a+a^*)$ and $a''=-\half i(a-a^*)$ are self-adjoint, such a map $\om'$ may be extended to a linear map
  $\om:B(H)\raw\C$ by $\om(a)=\om'(a')+i\om'(a'')$, which is positive in the equivalent sense that $\om(a^*a)\geq 0$ for each $a\in B(H)$, and still satisfies $\om(1_H)=1$. Thus $\om$ is a state in the modern sense, cf.\ \S\ref{sec1}.
   
Von Neumann's controversial argument against hidden variables is a simply corollary of his characterization of states, which we review first.  As already mentioned in \S\ref{sec1}, if $H$ is finite-dimensional, then any state $\om:B(H)\raw\C$ takes the form
 $\om(a)=\Tr(\rh a)$ for some density operator $\rh$ (i.e., $\rh\geq 0$ and $\Tr(\rh)=1$). In general, density operators define \emph{normal} states, which by definition have the following property:  for each orthogonal family $(e_i)$ of projections on $H$ (i.e., $e_i^*=e_i$ and  $e_ie_j=\dl_{ij}e_i$) one has $\om\left(\sum_i e_i\right)=\sum_i\om(e_i)$; states that are not normal are called \emph{singular}, and such states exist even if $H$ is separable, see below. 

Now suppose that some non-zero linear map $\om':B(H)_{\mathrm{sa}}\raw\R$ is \emph{dispersion-free}, i.e., $\om'(a^2)=\om'(a)^2$; without loss of generality we may also assume that $\om'(1_H)=1$. 
This implies that $\om'$ is positive and canonically extends to a state $\om:B(H)\raw\C$. If $\om'$ is also normal (which is automatic if $H$ is finite-dimensional), then  $\om(a)=\Tr(\rh a)$, as above, from which (provided $\dim(H)>1$)  it is easy to see that $\om$ cannot be dispersion-free on all self-adjoint operators $a$. In the eyes of von  Neumann and most of his contemporaries, this contradiction proved that \qm\ did not admit underlying hidden variables (at least of a kind we now  call \emph{non-contextual}). Belinfante (1973, p.\ 24, p. 34) recalls that
\begin{quotation}
 `the authority of von Neumann's overgeneralized claim for nearly two decades stifled any progress in the search for hidden-variable theories (\ldots) for decades nobody spoke up against von Neumann's arguments, and his conclusions were quoted by some as the gospel'. 
\end{quotation}
However, the  philosopher Hermann (1935)  already pointed out that von Neumann's linearity assumption, though valid for quantum-mechanical expectation values, was not justified
for dispersion-free states, if only because addition of non-commuting observables was not physically defined (a point of which von Neumann (1932) was actually well aware, since he attempted to justify such additions through the use of ensembles). 
See also Seevinck (2012). Unaware of this---Hermann   was one of the very few women in academia at the time and also published in an obscure Neokantian journal---Bell (1966) and Kochen \& Specker (1967) made a similar point and also remedied it, proving what we now call the Kochen--Specker Theorem (sometimes also named after Bell).
It is sad to note that  Bell and some of his followers later  left the realm of decent academic discourse (and displayed the depth of their own misunderstanding) by calling von Neumann's argument ``silly'' and ``foolish''.
 In  fact,  von Neumann  carefully qualifies his  result by stating that it follows
  `\emph{im Rahmen unserer Bedingungen}', i.e.\ `\emph{given our assumptions}', and 
 should be credited rather than ridiculed for being the first author to impose useful constraints on \hv\ theories, anticipating  all later literature on the subject; see also  Bub (2011).

 In order to explain the link of this development to (exact) Bohrification, we now go back to Gleason (1957), but in addition use later results as reviewed in Maeda (1990) and Hamhalter (2004). We first collect all relevant notions in to a single definition. 
 \begin{Definition}\label{gleasondefs2}
Let $H$ be an arbitrary \Hs\ with projection lattice 
\beq\CP(H)=\{e\in B(H)\mid e^2=e^*=e\}.
\eeq
\begin{enumerate}
\item A map $P:\CP(H)\raw[0,1]$ that satisfies  $P(1_H)=1$ is called:
\begin{enumerate}
\item  a \emph{finitely additive probability measure} if 
\beq
P\left(\sum_{j\in J} e_j\right)=\sum_{j\in J}P(e_j) \label{whatsum}
\eeq for any \emph{finite} 
collection $(e_j)_{j\in J}$ of mutually orthogonal projections on $H$
(i.e., $e_jH\perp e_k H$, or equivalently, $e_je_k=0$, whenever $j\neq k$);   equivalently,
$$ P(e+f)=P(e)+P(f) \mbox{ whenever } ef=0.$$
\item a \emph{probability measure} if \er{whatsum} holds for any \emph{countable} collection $(e_j)_{j\in J}$ of mutually orthogonal projections on $H$;
\item  a \emph{completely additive probability measure} if  \er{whatsum} holds for  \emph{arbitrary} 
collections $(e_j)_{j\in J}$ of mutually orthogonal projections on $H$.
\end{enumerate}
In (b) and (c) the sum $\sum_j$ is defined in the \emph{strong} operator topology on $B(H)$.
\item A \emph{strong quasi-state} on $B(H)$ is a map $\om:B(H)\raw\C$ that is positive (that is, $\om(a^*a)\geq 0$ for each $a\in B(H)$) 
and normalized (i.e.\ $\om(1_H)=1$), and otherwise:
\begin{enumerate}
\item satisfies $\om(a)=\om(a')+i\om(a'')$, where $a'= \half (a+a^*)$ and $a''=-\half i(a-a^*)$.
\item is linear on all commutative unital \ca s in $B(H)$. 
\end{enumerate}
Note that $a'$ and $a''$ are self-adjoint, so that $\om$ is fixed by its values on $B(H)_{\mathrm{sa}}$. 
Hence we have $\om(z a)=z\om(a)$, $z\in\C$, and $\om(a+b)=\om(a)+\om(b)$ whenever $ab=ba$.
\item  A \emph{weak quasi-state} on $B(H)$ satisfies all properties of a strong one except that linearity is only required on   \emph{singly generated} commutative \ca s in $B(H)$, i.e., those of the form $C^*(a)$, where $a=a^*\in B(H)$; this  is the  \ca\ generated by $a$ and $1_H$ (which is the norm-closure of the space of all finite polynomials in $a$).
\item A \emph{strong (weak) dispersion-free state}  on $B(H)$ is a strong (weak) quasi-state  that is pure on each (singly generated) commutative unital \ca\ in $B(H)$.
\end{enumerate}
\end{Definition}
It was shown by Aarens (1970) that the map $\om\mapsto \om_{|\CP(H)}$ gives a bijective correspondence between weak quasi-states $\om$ on $B(H)$ and
finitely additive probability measures on $\CP(H)$.

A probability measure is by definition $\sg$-additive in the usual sense of measure theory; the other two cases are unusual from that perspective. However, if $H$ is separable, then $J$ can be at most countable, so that complete
additivity is the same as $\sg$-additivity and hence any probability measure is automatically completely additive.
Surprisingly, assuming the \emph{Continuum Hypothesis} (CH) of set theory, it can be shown that this is even the case for arbitrary \Hs s  (Eilers \& Horst, 1975). The fundamental distinction, then, is between 
\emph{finitely} additive probability measures and probability measures (which by definition are \emph{countably} additive). As we shall see shortly, this reflects the (often neglected) distinction between \emph{arbitrary} and \emph{normal} states on $B(H)$, respectively. Indeed, we now have the following generalization (and bifurcation) of Gleason's Theorem:
\begin{Theorem}\label{Gleason2}
Let $H$ be a  \Hs\ of dimension greater than 2.
\begin{enumerate}
\item Each probability measure $P$ 
on $\CP(H)$ is induced by a   normal state  on $B(H)$ via
\beq
P(e)=\Tr(\rh e), \label{Peometrrhoe}
\eeq where $\rh$ is a  density operator on $H$ uniquely determined by $P$. 
 Conversely, each density operator $\rh$ on $H$ defines a probability measure $P$ 
on $\CP(H)$ via \er{Peometrrhoe}. 
Without CH, this is true  also for non-separable $H$ if $P$ is assumed to be completely additive. 
\item  Each finitely additive probability measure $P$ 
on $\CP(H)$ is induced by a  unique  state  $\om$ on $B(H)$ via
\beq
P(e)=\om(e).\label{Peometrrhoec}
\eeq 
 Conversely, each state $\om$ on $H$ defines a probability measure $P$ 
on $\CP(H)$ via \er{Peometrrhoec}.
\end{enumerate}
\end{Theorem}
\begin{Corollary} If $\dim(H)>2$, then
 each weak quasi-state on $B(H)$  is linear and hence is actually a state. In particular, 
 weak and strong quasi-states coincide.
\end{Corollary}
In this language, the Kochen--Specker Theorem takes the following form:
 \begin{Theorem}\label{KSv1}
 If $\dim(H)>2$,  there are no strong (weak) quasi-states $\om:B(H)\raw\C$ whose restriction to each (singly generated) 
 commutative C*-subalgebra of  $B(H)$ is pure.
 \end{Theorem}
 For an efficient proof see  D\"{o}ring (2005). In summary, the (trivial) first step is to show that
 $\om(e)\in\{0,1\}$ for any projection $e$. From this, one shows that $\om$ must be multiplicative on all of $B(H)$ (Hamhalter, 1993). If $\om$ is normal, Gleason's Theorem already excludes this. If not, $H$ must be infinite-dimensional, in which case 
  there is a projection $e$ and an operator $v$ such that $e=vv^*$ and $1_H-e=v^*v$ (this is sometimes called the \emph{halving lemma}, which is easy to prove).  Multiplicativity of $\om$ then implies   $\om(e)=\om(1_H-e)$, which 
contradicts the additivity property $\om(e)+\om(1_H-e)=\om(1_H)=1$ obtained from Aarens (1970) cited above: indeed, 
if $\om(e)=0$ one finds $0=1$, whereas $\om(e)=1$ implies $2=1$. 
 
  To understand the connection with the original 
  Kochen--Specker Theorem, let some map $\om':B(H)_{\mathrm{sa}}\raw\R$ be dispersion-free, i.e., $\om'(a^2)=\om'(a)^2$ for each 
 $a\in B(H)_{\mathrm{sa}}$, and quasi-linear, i.e., linear on commuting operators (it was shown by Fine (1974) that these conditions are equivalent to the ones originally imposed by Kochen \& Specker (1967), viz.\ $\om(a)\in\sg(a)$ and
 $\om(f(a))=f(\om(a))$ for each Borel function $f:\sg(a)\raw\R$).
 Canonically extending such a map $\om'$ to a map $\om:B(H)\raw\C$ by complex linearity 
 (cf.\ Definition \ref{gleasondefs2}.2 (a)) 
  and noting that
 dispersion-freeness implies positivity and hence continuity on each subalgebra $C^*(a)$, we see that 
  dispersion-freeness and quasi-linearity
  imply that $\om$ is multiplicative on $C^*(a)$, and hence pure (conversely, pure states on $C^*(a)$ are dispersion-free). Finally, although linearity on all commuting self-adjoint operators seems stronger than linearity on each $C^*(a)$, Theorem \ref{Gleason2} shows that these conditions are in fact equivalent. 
  
  In sum,  Gleason's Theorem and the Kochen--Specker Theorem  give results (of a positive and a negative kind, respectively) on the behaviour of state-like functionals on all of $B(H)$ \emph{given their behaviour on commutative C*-subalgebras thereof}. This supports the idea of exact Bohrification, which suggests that access to $B(H)$ should precisely be gained through such subalgebras. We now turn to another major influence on this kind of thinking.
\section{The Kadison--Singer conjecture}\label{KSC}
A state $\om$ on some \ca\ $A$ is called \emph{pure} if it has no decomposition
$\om=t\om_1+(1-t)\om_2$, where $t\in(0,1)$ and $\om_1\neq\om_1$ are states on $A$ (note that the set of all states is convex). The pure states on $A=C_0(X)$ are given by the evaluation maps $\om_x(f)=f(x)$ and hence correspond to points $x\in X$. For any \Hs\ $H$, the \emph{normal} pure states on $B(H)$ are the vector states $\om_{\ps}(b)=\la\psi,b\psi\ra$, where $\psi\in H$ is some unit vector.

Around the same time as Gleason (1957), and similarly inspired by \qm\ through the work of von Neumann (1932) and Mackey (1957), Kadison \& Singer (1959) wrote a visionary paper on pure states on \emph{maximal} commutative C*-subalgebras of $B(H)$. 
Their work   arose from the  desire to clarify a potential mathematical ambiguity in the Dirac notation commonly used in \qm.
Namely, if $\ul{a}=(a_1, \ldots, a_n)$ is
 a maximal set of commuting self-adjoint operators on some \emph{finite-dimensional} \Hs\ $H$,   
then $H$ has a basis of joint eigenvectors $|\ul{\lm}\ra$ of $\ul{a}$, which physicists typically label by the corresponding eigenvalues 
$\ul{\lm}=(\lm_1, \ldots, \lm_n)$. Equivalently, given a single  self-adjoint operator $a$ that is maximal in the sense that its spectrum $\sg(a)$ is nondegenerate, the eigenvectors of $a$, denoted by $|\lm\ra$, $\lm\in\sg(a)$, form a basis of $H$. The first situation reduces to the second, since $a_i=f_i(a)$ for a single $a$ and suitable functions $f_i:\sg(a)\raw\R$. 

Algebraically, what happens here is that each $\lm\in\sg(a)$ initially  defines a pure (and hence multiplicative) state $\om'_{\lm}$ on $C^*(a)$ by  $\om'_{\lm}(a)=\lm$ and of course  $\om'_{\lm}(1_H)=1$, which enforces
 $\om'_{\lm}(f(a))=f(\lm)$ for any polynomial $f$ (this much is true also if $\dim(H)=\infty$, in which case the last equation holds  for any $f\in C(\sg(a))$, see also \S\ref{Born} below).  Since $\lm$ defines a unit eigenvector $|\lm\ra$ that, given that $a$ is maximal, is unique up to a phase, it also defines a unique  extension $\om_{\lm}$ of $\om'_{\lm}$  to a pure state  on  $B(H)$, namely (in Dirac notation) $\om_{\lm}(b)=\la\lm|b|\lm\ra$, $b\in B(H)$. Indeed, it is this uniqueness property that
 makes the labeling $|\lm\ra$ unambiguous in the first place. By contrast, if $a$ is not maximal  it has an  eigenvalue $\lm$ having at least two orthogonal eigenvectors, which clearly define different vector states on $B(H)$. Similarly for 
 the commutative \ca\   $C^*(\ul{a})$ generated by the $a_i$;  the two examples can be united by noting that---assuming maximality---both $C^*(a)$ and $C^*(\ul{a})$ are unitarily equivalent to the algebra $D_n(\C)$ of diagonal matrices on $\C^n$ for  $n=\dim(H)$ (namely by changing some given basis of $H$  to a basis of eigenvectors of  $a$ or $\ul{a}$). 
  
 The same is true for any maximal commutative \ca\ $A\subset B(H)$, since $A=C^*(a)$ for some maximal self-adjoint $a$.
We may therefore summarize the discussion so far as follows: \emph{if $H$ is finite-dimensional, then
  any pure state on any maximal commutative \ca\ $A\subset B(H)$ has a unique extension to a pure state on $B(H)$, and this is what makes  the Dirac notation $|\lm\ra$ unambiguous.} This supports the idea of exact Bohrification,  since it shows that a state on the noncommutative algebra  $B(H)$ of all observables  is already determined by its value on any maximal commutative C*-subalgebra $A\subset B(H)$.
 
 What about infinite-dimensional \Hs s? The first difference with the finite-dimensional case is that even if $H$ is separable (which is the only case that has been studied so far, sufficient as it is for most applications), maximal commutative C*-subalgebras $A$ of $B(H)$---which are always \vna s, i.e.\ $A''=A$---are no longer unique (up to unitary equivalence). Indeed,  $A$
is unitarily equivalent to exactly one of the following:
\begin{enumerate}
\item $L^{\infty}(0,1)\subset B(L^2(0,1))$;
\item $\ell^{\infty}(\N)\subset B(\ell^2(\N))$;
\item $L^{\infty}(0,1)\oplus \ell^{\infty}(\kappa)\subset B(L^2(0,1)\oplus\ell^2(\kappa))$,
\end{enumerate}
where $\kappa$ is either $\{1,\ldots, n\}$, 
in which case $\ell^2(\kappa)\cong\C^n$ and $\ell^{\infty}(\kappa)\cong D_n(\C)$, or $\kp=\N$;
the inclusions are given by realizing each commutative algebra by multiplication operators. This classification was stated without proof in Kadison \& Singer (1959); the details appeared later in  Kadison \& Ringrose (1986, \S9.4), based on von Neumann (1931), who initiated the study of commutative \vna s. See also Stevens (2015). 

 The second difference  is that if $\dim(H)=\infty$, then $B(H)$ admits \emph{singular} pure states, which cannot be represented by some unit vector in $H$; the difficulty of the state extension property in question entirely comes from the singular case. 
We elaborate a little. If $H=L^2(0,1)$ and $a$ is the position operator, i.e., $a\psi(x)=x\psi(x)$, having spectrum $\sg(a)=[0,1]$, then for any $\lm\in\sg(a)$ there exists a (necessarily singular) pure state $\om$ on $B(H)$ such that $\om(a)=\lm$. More generally, this is true whenever $a\in B(H)_{\mathrm{sa}}$ and $\lm\in\sg_c(a)$ (i.e., the continuous spectrum of $a$, defined as the complement of the set of eigenvalues in $\sg(a)$). 

This apparently gives rigorous meaning to the Dirac notation $|\lm\ra$ for improper eigenstates, though with an important caveat. In the light of the discussion above, one might rephrase the  situation as follows: the position operator $a$ generates a maximal commutative \ca\ $L^{\infty}(0,1)$ in $B(L^2(0,1))$, which admits no \emph{normal} pure states whatsoever. However, each $\lm\in\sg(a)$ defines a \emph{singular} pure state  $\om'_{\lm}$ on $L^{\infty}(0,1)$, which has at least one pure extension $\om_{\lm}$ to $B(L^2(0,1))$. But is this extension unique? This question also arises for 
$\ell^{\infty}(\N)$, although this algebra does admit normal pure states (namely $\om_k(f)=f(k)$, where $k\in\N$). However, the ``overwhelming majority'' of pure states on $\ell^{\infty}(\N)$ are singular (corresponding to elements of the so-called  \v{C}ech--Stone compactification $\beta\N$ of $\N$). 

  For separable \Hs s, the pure state extension problem has been  solved:
  \begin{Theorem}\label{KSTheorem12}
\begin{itemize}
\item Any \emph{normal} pure state on any maximal commutative C*-subalgebra $A\subset B(H)$ has a unique (and hence pure) extension to $B(H)$.
\item Any pure state on $\ell^{\infty}(\N)$ has a unique extension to $ B(\ell^2(\N))$.
\item There exist (necessarily singular) pure states on $L^{\infty}(0,1)$ that do not have a unique extension to $B(L^2(0,1))$, and similarly for $L^{\infty}(0,1)\oplus \ell^{\infty}(\kappa)$, for any $\kappa$. 
\end{itemize} 
\end{Theorem}
We note 
that uniqueness of the the state extension property in question---or the lack of it---is preserved under unitary equivalence.
The first part of this theorem is even true for non-separable \Hs s; it may be proved along the lines of the finite-dimensional case. If one agrees that only normal states (i.e., density operators) can be realized physically---in that singular states (such as exact eigenstates of position or momentum) are idealizations---this settles the issue and supports exact Bohrification, as explained above. The singular case, however, is of much more mathematical interest and might even be physically relevant, depending on the way one thinks about idealizations (see also \S\ref{ideal} below).

 The second part of the theorem is the path-breaking solution of the \emph{Kadison--Singer conjecture} due to Marcus, Spielman,  \& Srivastava (2014a,b), with important earlier contributions by Weaver (2004);  see also Tao (2013) and Stevens (2015) for lucid expositions of the proof. Though less well known, the third part, due to Kadison \& Singer themselves, whose arguments were later simplified by Anderson (1979) and  Stevens (2015), is as  remarkable than the second; it shows that Dirac's notation $|\lm\ra$ may be ambiguous, or, equivalently, that  maximal commutative C*-subalgebras of $B(H)$ that are unitarily equivalent to $L^{\infty}(0,1)$ (like the one generated by the position operator or the momentum operator) do not suffice to characterize pure states. What to make of this is unclear. 
  \section{The poset $\CA$, topos theory, and quantum logic}
  Returning to the story in \S\ref{GKS}, the next step was to collect all unital  commutative C*-subalgebras of $B(H)$, or indeed of an arbitrary unital
   \ca\ $A$, into a single mathematical object $\CA$; with some goodwill, one might call $\CA$ the mathematical home of complementarity (although the construction applies even when $A$ itself is commutative). This decisive idea goes back to  Isham \& Butterfield (1998) for $A=B(H)$ and, more generally, to Hamilton, Isham, \& Butterfield (2000) for arbitrary \vna s (and hence these authors consider  von Neumann subalgebras instead of C*-subalgebras).
   
Heunen, Landsman \& Spitters (2009) introduced the poset $\CA$ whose elements are  commutative C*-subalgebras of $A$ that contain the unit $1_A$, ordered by (set-theoretic) inclusion (note that Isham et al use the opposite order, which even apart from their von Neumann algebraic setting gives the theory a totally different flavor as soon as sheaves are defined).
  At this stage of the Bohrification program it was perhaps not  realized what a powerful object $\CA$ already  is by itself, even though it has ``forgotten'' the commutative subalgebras (which are merely points in $\CA$) and just ``remembers'' their inclusion relations; this realization only came with papers like D\"{o}ring \& Harding (2010),  Hamhalter (2011), Heunen (2014a),
    Heunen \& Lindenhovius (2015), and  Lindenhovius (2016). 
      Thus we introduced the ``tautological map'' $C\mapsto C$, where the $C$ on the left-hand side is an element of $\CA$ (now seen as a posetal category, in which $C$ and $D$ are connected by an arrow iff $C\subseteq D$), whereas the $C$ on the right-hand side is $C$ itself, initially seen as a set. Seen as a functor $\ul{A}$ from $\CA$ to the category $\mathsf{Sets}$ of sets (where the arrow $C\raw D$, i.e., $C\subseteq D$, is mapped to the inclusion map $C\hookrightarrow D$ of the underlying sets), 
our tautological map turns out to define a \emph{commutative} internal \ca\ in the topos $[\CA,\mathsf{Sets}]$
 of all functors from $\CA$ to $\mathsf{Sets}$. See Mac Lane \& Moerdijk (1994) for an introduction to topos theory and Banaschewski \& Mulvey (2006) for the general theory of commutative \ca s and Gelfand duality in toposes: broadly speaking, a topos provides a universe---alternative to set theory---in which to do mathematics, generally based on intuitionistic rather than classical logic, see below. It was this transfiguration of a (generally noncommutative) \ca\ $A$ into an internal commutative \ca\ $A$, which essentially consists of all (unital) commutative C*-subalgebras of $A$ but may be treated as a single \ca\ (in a different topos from $\mathsf{Sets}$),  that was initially meant by the term ``Bohrification'', which we now use much more liberally. So far, the main harvest of this construction has been:
 \begin{itemize}
\item The definition (Heunen et al, 2009) and explicit computation (Heunen et al, 2012; Wolters, 2013) of the (internal) Gelfand spectrum $\ul{\Sg}(\ul{A})$ of $\ul{A}$,  playing the role of a quantum-mechanical phase space (which is lacking in the usual formalism).
\item The realization of states on $A$ as probability measures on  $\ul{\Sg}(\ul{A})$  (Heunen et al, 2009).
\item The reformulation of the Kochen--Specker Theorem to the effect that (at least for \ca s like $A=B(H)$ for $\dim(H)>2$) 
the phase space $\ul{\Sg}(\ul{A})$ has no points (Isham \& Butterfield, 1998; Caspers et al, 2009; Heunen et al, 2009).
\item Intuitionistic logic as the logic of \qm\ (Caspers et al, 2009;  D\"{o}ring \& Isham, 2010; 
D\"{o}ring 2012; Heunen et al, 2012; Hermens, 2009, 2016; Wolters, 2013).
\end{itemize}
All this requires extensive explanation, especially because these claims are meant in the sense of internal reasoning in a topos (Mac Lane \& Moerdijk, 1994), but let us restrict ourselves to some brief comments on the last two points (see references for details).

 Even in the topos $\mathsf{Sets}$, which is the home of classical mathematics (using the Zermelo--Fraenkel axioms), spaces may or may not have points, provided we interpret the notion of a ``space'' as a \emph{frame}, which is a (necessarily distributive)
complete  lattice $F$ in which $U\wedge \bigvee S=\bigvee \{U\wedge V, V\in S\}$
 for arbitrary elements $U\in F$ and subsets $S\subset F$. Frames are motivated by the example $F=\CO(X)$,
 i.e., the collection of open sets on some topological space $X$, where the supremum is  given by  set-theoretic union.
 To see if a given frame $F$ is of this form, one looks at the \emph{points} of $F$, defined as frame maps $p:F\raw\{0,1\}$, where
 the frame $\{0,1\}\equiv \CO(\mathrm{point})$ has order $0\leq 1$. The set $\mathrm{Pt}(F)$ of points of $F$ is a topological space, with open sets $\{p\in \mathrm{Pt}(F)\mid p\inv(U)=1\}$, where $U\in F$. If $F\cong \CO(\mathrm{Pt}(F))$, we say that $F$ is \emph{spatial}.  Even in set theory, frames are not necessarily spatial (and if they are, the axiom of choice---which is unavailable in many toposes---is often needed to prove this).
 All this can be internalized in topos theory, and the Kochen--Specker Theorem states that the Gelfand spectrum of $\ul{\Sg}(\ul{A})$ of $\ul{A}$ not merely fails to be spatial: it has no points whatsoever! Looking at points of phase space in classical physics as truth-makers (of propositions), this is no surprise: after all,  the very point (sic) of the Kochen--Specker Theorem is  that in \qm\ not all propositions can have simultaneous (sharp) truth values. 
 
 Frames are also closely related to \emph{Heyting algebras}, defined as distributive lattices ${H}$ (with top $1$ and bottom $0$) equipped with a binary map $\raw:{H}\x{H}\raw{H}$ (playing the role of implication in logic) that satisfies the axiom 
$U\leq (V\raw W)$   iff  $(U\wed V)\leq W$. In a Heyting algebra (unlike a Boolean algebra), negation is a derived
notion, defined by $\neg U=U\raw\bot$. 
 Every Boolean algebra is
a Heyting algebra, but not \emph{vice versa}; in fact, a Heyting
algebra is Boolean iff $\neg\neg U=U$ for all $U$, or, equivalently,  $(\neg U)\vee U=\top$,
which states the law of the excluded middle (famously denied by the Dutch mathematician L.E.J. Brouwer). 
Thus Heyting algebras formalize intuitionistic propositional logic. 
A  Heyting algebra  $H$ is \emph{complete} when it is complete as a lattice, in which case $H$ is a frame.
 Conversely,  a frame is a complete Heyting algebra with implication $V\raw W=\bigvee\{U\mid U\wed V\leq W\}$. 
 
The point, then, is that exact Bohrification implies that quantum logic is given by a particular Heyting algebra $H(A)$ eventually  defined by the given \ca\ $A$.  Thus quantum logic turns out to be intuitionistic: it preserves distributivity (i.e., of ``and'' over ``or'') but denies the law of the excluded middle. The simplest example is $A=M_n(\C)$, which describes an $n$-level system, for which the relevant Heyting algebra comes down as
\begin{equation}
H(M_n(\C))=\{\mathsf{e}:\mathcal{C}(A) \raw \CP(A)\mid {\mathsf{e}}(C)\in \CP(C),\, 
{\mathsf{e}}(C)\leq {\mathsf{e}}(D)\:\mbox{if}\: C\subseteq D\}, \label{bohr}
\end{equation}
with pointwise order, i.e., $\mathsf{e}\leq \mathsf{f}$ iff $\mathsf{e}(C)\leq \mathsf{f}(C)$ for each $C\in\C(M_n(\C))$. We see that whereas in the traditional (von Neumann) approach to quantum logic a \emph{single} projection $e\in\CP(A)$ defines a proposition, Bohr's doctrine of classical concepts suggests that a proposition consists of a \emph{family}  $\mathsf{e}$ of projections $\mathsf{e}(C)$, one  for each classical context $C$.

 Intuitionistic logic goes flatly against Birkhoff \& von Neumann (1936), whose quantum logic has exactly the opposite features in denying distributivity but keeping the law of the excluded middle. It also seems to go against Bohr, who in 1958 claimed that
\begin{quotation}
`all departures from common language and ordinary logic are entirely avoided by reserving the word ``phenomenon'' solely for reference to unambiguously communicable information, in the account of which the word ``measurement'' is used in its plain meaning of standardized comparison.' (Bohr, 1996, p.\ 393)
\end{quotation}
However,  the preceding text  makes it clear that Bohr refers to certain multivalued logics (Gottwald, 2015) rather than to the intuitionistic logic of Brouwer and Heyting, which might even model everyday reasoning  better than classical logic does (Moschovakis, 2015).
  \section{The Born rule revisited}\label{Born}
  In this section we apply exact Bohrification to singly generated  commutative \ca s in $A=B(H)$, where $H$ is an arbitrary \Hs\ (see \S\ref{GKS}), so let  $a=a^*\in B(H)$ and consider the \ca\ $C^*(a)$ generated by $a$ and $1_H$. 
   Another commutative \ca\ defined by $a$ is $ C(\sg(a))$, i.e., the algebra of continuous complex-valued functions on the spectrum $\sg(a)$ of $a$ (which is a compact subset of $\R$), equipped with pointwise addition, (scalar)  multiplication, and complex conjugation (as the involution), and the supremum norm, cf.\ \S\ref{sec1}. 
  One version of the spectral theorem now states that $C(\sg(a))$ and $C^*(a)$ are isomorphic (as
  commutative  \ca s) under a map $f\mapsto f(a)$ that  is given by (norm-) continuous extension of the map
  defined on polynomial functions simply by $p\mapsto p(a)$, that is, if $p(x)=
\sum_n c_n x^n$,  where $x\in \sg(a)$ and $c_n\in \C$, then $p(a)= \sum c_n a^n$. In particular, 
 the unit function $1_{\sg(a)}:x\mapsto 1$ maps to  the unit operator $1_{\sg(a)}(a)=1_H$, whilst 
 the identity function $\mathrm{id}_{\sg(a)}:x\mapsto x$ is mapped to  $\mathrm{id}_{\sg(a)}(a)=a$. See e.g.\ Pedersen (1989), Thm.\ 4.4.1. 
 
Now recall the Riesz--Radon representation theorem from \S\ref{sec1}.
Take a unit vector $\psi\in H$ (or, more generally,   a state $\om$ on $B(H)$). Combining this with the spectral theorem  above, we obtain a  unique probability measure $\mu_{\psi}$ (or $\mu_{\om}$)
 on the spectrum $\sg(a)$ of $a$ such that
 \beq
 \la\ps, f(a)\ps\ra=\int_{\sg(a)} d\mu_{\ps}\, f,
 \eeq for each $f\in C(\sg(a))$ (or, more generally, $\om(f(a))=\int_{\sg(a)} d\mu_{\om}\, f$). This is the Born measure; for example, if $\dim(H)<\infty$, we recover the familiar result 
 \beq
 \mu_{\psi}(\{\lm\})=\| e_{\lm}\ps\|^2, \label{Born1}
 \eeq where $e_{\lm}$ is the spectral projection onto the eigenspace $H_{\lm}\subset H$ for the eigenvalue $\lm\in\sg(a)$; of course, the right-hand side equals $\la\ps,e_{\lm}\ps\ra$. 
 If the spectrum is non-degenerate, this expression becomes $|\la\ups_{\lm},\ps\ra|^2$, where $\ups_{\lm}$ is ``the'' unit eigenvector of $a$ with eigenvalue $\lm$. 
  
Although the underlying mathematical reasoning is well known (Pedersen, 1989, \S4.5), the thrust of (exact) Bohrification is 
that the Born rule---which is the main connection between the mathematical formalism of \qm\ and laboratory physics---comes from a simple look at appropriate commutative subalgebras of $B(H)$. Namely, the (vector) state $b\mapsto \la\ps, b\ps\ra$ on $B(H)$ defined by $\psi\in H$ (or indeed any state on $B(H)$) is simply restricted to $C^*(a)\subset B(H)$, and is subsequently transferred to $C(\sg(a))$ through the spectral theorem, where it becomes a probability measure (Landsman, 2009).
 One might argue that this argument validates Heisenberg's (1958, pp.\ 53--54) claim  that 
\begin{quotation} `One may call these uncertainties objective, in that they are simply a consequence of the fact that we describe the experiment in terms of classical physics; they do not depend in detail on the observer. 
One may call them subjective, in that they reflect our incomplete knowledge of the world.'
\end{quotation}
Nonetheless, by itself such mathematical reasoning is insufficent to \emph{derive} the Born rule, but it does show that all one needs to postulate in \qm\ in this respect is  its \emph{function} rather than its \emph{form} (which, as shown above, is simply given by the formalism). 

Asymptotic Bohrification closes also this gap. Following Landsman (2008) we
complete (and---in the light of valid critique by  Cassinello \& S\'{a}nchez-G\'{o}mez (1996) and Caves \& Schack
 (2005)---correct) a program begun by Finkelstein (1965) and Hartle (1968), as further developed by Farhi, Goldstone, \& Gutmann (1989), and Van Wezep (2006).

Referring to the appendix for notation and background, let  $I=1/\dot{\N}$, take some fixed unital \ca\ $B$ (for which we will just need the $n\x n$ matrices $B=M_n(\C)$), and define a bundle $(A_{\hbar})_{{\hbar}\in I}$ of \ca s by taking
$A_{1/N}=B^N$ (i.e., the $N$-fold projective tensor product $\ot^N B$ of $B$ with itself) and $A_0=C(S(B))$, where $S(B)$ is the state space of $B$, seen as a compact convex set in the weak$\mbox{}^*$-topology. For example, the state space of $B=M_2(\C)$ is affinely homeomorphic to the unit ball in $\R^3$, whose boundary is the familiar Bloch sphere of qubits.
As we shall see, $A_0$  is going to be the commutative \ca\ through which we look at the (typically non-commutative) \ca\ $B$. To this effect, 
 we now turn this family into a continuous bundle of \ca s by stipulating what the continuous  sections are (Landsman, 2007, 2008). We  first define the usual
symmetrization operator $S_N: B^N\raw B^N$ by linear (and if necessary continuous) extension of
 \beq
S_N(b_1\ot\cdots \ot b_N)=\frac{1}{N!}\sum_{\sg\in \GS_N} b_{\sg(1)}\ot\cdots\ot b_{\sg(N)}, \label{landc}
\eeq
where $\GS_N$ is the permutation group (i.e.\ symmetric group) on $N$ elements, and $b_i\in B$ for all $i=1,\ldots,N$. 
For $N\geq M$ we then define $j_{NM}: B^M\raw B^M$ by linear extension of
\beq
j_{NM}(a_M)=S_N(a_M\ot 1_B\ot\cdots \ot 1_B), \label{symmaps}
\eeq 
where one has $N-M$ copies of the unit $1_B\in B$ so as to obtain an element of $B^N$.

For example, the symmetrizer $j_{N1}:B\raw B^N$ is given by
\beq
j_{N1}(b)=\frac{1}{N}\sum_{k=1}^N 1_B\otimes\cdots\ot b_{(k)}\ot 1_B\cdots \otimes 1_B,\label{j1N}\eeq
which is just the ``average'' of $b$ over all $N$ copies of $B$. Our main examples  of \er{j1N} will be  \emph{frequency operators},
where $b=e$ for  some projection $e\in B$ (i.e., $e^2=e^*=e$). In particular, if $a=a^*\in B =M_n(\C)$ and $\lm\in\sg(a)$, we may take the spectral projection $e_{\lm}$. Applied to states of the kind $\ups_1\ot\cdots\ot\ups_n\in\C^N$, where each
$\ups_i$ is an eigenstate of $a$, so that $a\ups_i=\lm_i\ups_i$ for some $\lm_i\in\sg(a)$, the corresponding operator \beq
f_N^{\lm}=j_{N1}(e_{\lm}) \label{Fofe}
\eeq counts the relative frequency of $\lm$ in the list $(\lm_1,\ldots, \lm_n)$. We return to the construction of a continuous bundle of \ca s with fibers $A_0$ and $A_{1/N}$, $N\in\N$. As explained in the appendix, given the fibers we may define the 
topological structure by specifying the continuous cross-sections. To this end, we say that a sequence $(a_N)_{N\in\N}$, with $a_N\in B^N$, is \emph{symmetric} when $a_N=j_{NM}(a_M)$
 for some fixed $M$ and all $N\geq M$, and \emph{approximately symmetric} if for any $\varep>0$ there is 
  a symmetric sequence $(a'_N)$ and some
  $N(\varep)\in\N$  such that 
$\|a_N-a_N'\|< \varep$ for all $N\geq N(\varep)$ (norm in $A_N$).  Each section $a$ of the bundle $(A_{\hbar})_{{\hbar}\in 1/\dot{\N}}$ is a sequence $(a_0, a_N)_{N\in\N}$, where $a_0\in C(S(B))$ and $a_N\in B^N$.  
The \emph{continuous} sections are those for which $(a_N)_{N\in\N}$ is approximately symmetric and $a_0$ is given by
 \beq
 a_0(\om)=\lim_{N\raw\infty}\om^N(a_N).\label{a0}
 \eeq  
 Here $\om\in S(B)$, and $\om^N\in S(B^N)$ is defined by linear (and continuous) extension of 
\beq \om^N(b_1\ot\cdots\ot b_N)=\om(b_1)\cdots\om(b_N);\label{omN}\eeq
 this limit exists by definition of an approximately symmetric sequence.
 The point of  this is that since the frequency operator \er{Fofe}, seen as a sequence $(f_N^{\lm})_{N\in\N}$, is  evidently  symmetric (and hence \emph{a fortiori} approximately symmetric), it therefore defines a continuous section of our bundle $(A_{\hbar})_{{\hbar}\in 1/\dot{\N}}$ if we complete it with its limit \er{a0}, which is given by
 \beq
 f_0^{\lm}(\om)=\om(e_{\lm}).
\eeq
This is  the Born probability for the outcome $a=\lm$ in the state $\om$; for example, if $B=M_n(\C)$ and
$\om(b)=\la\ps,b\ps\ra$ for some unit vector $\psi\in\C^n$, then $\om(e_{\lm})=\|e_{\lm}\ps\|^2$, cf.\ \er{Born1}.

The physical interpretation of this mathematical result is as follows: if the system is prepared in some state $\om$, 
the number $f_0^{\lm}(\om)$ is (by construction) equal to the limiting frequency of either a single experiment on a large number of sites or a long run of individual experiments on a single site. A classical perspective on either of those \emph{a priori} quantum-mechanical situations only arises in the limit $N\raw\infty$ (i.e., $1/N\raw 0$), in which  the Born probability arises as an objective property of the experiment(s).  Nothing is implied about single cases; these will be discussed in \S\ref{mmp} below, preceded by an intermezzo on the complications caused by the fact that in reality the limit $N=\infty$ (or $\hbar=0$) is never reached. This is arguably why Bohr insists that merely the \emph{results of the observations}---as opposed to the underlying \emph{phenomena}---must be expressed in classical terms. 

 Our two derivations of the Born rule reflect the two ways one may look at probabilities in physics (as opposed to betting, which context is inappropriate here), namely either as relative frequencies (which corresponds to the point of view offered by asymptotic  Bohrification) or as chances for outcomes of individual random events (which is the perspective given by exact  Bohrification); the relationship between these two approaches to probability is as much in need of clarification as  the link between exact and asymptotic Bohrification!
\section{Intermezzo: Limits and idealizations}\label{ideal}
Asymptotic  Bohrification always involves some classical theory, described by a commutative \ca\ of observables $A_0$ (typically of the form $A_0=C_0(X)$ for some phase space $X$) and a family of quantum theories indexed by $\hbar\in I\subset [0,1]$, each of which is described by some commutative \ca\ of observables $A_{\hbar}$. This  includes both the case where 
$\hbar\in(0,1]$ and the limit of interest is $\hbar\raw 0$, and the case where
$\hbar=1/N$, where $N\in\N$  and the limit of interest is $N\raw\infty$. We regard the $A_0$-theory as an idealization that never occurs in physical reality, but which  approximates the family of $A_{\hbar}$-theories in asymptotic regimes described by the parameter $\hbar$ (or $N$). Thus we follow Norton (2012, abstract):
\begin{quotation} [$A_0$ is]
`another system whose properties provide an inexact description of the target system' [i.e., $A_{\hbar}$].
\end{quotation}
Mathematically, $A_0$  is a limiting case of the theories  $A_{\hbar}$
as $\hbar\raw 0$ (or $N\raw\infty$), in the sense explained in the appendix. 
Examples one may have in mind are:
\begin{itemize}
\item  $A_0$ is either classical mechanics or thermodynamics (of an infinite system);
\item $A_{\hbar}$ is either quantum mechanics or quantum statistical mechanics (for finite $N$).
\end{itemize}
See Landsman (1998) for the classical-quantum link in this language and  Landsman (2007) for thermodynamics as a limiting case of quantum statistical mechanics of finite systems.

\noindent 
The latter also has another limit, namely quantum statistical mechanics of infinite systems, described by a different continuous bundle of \ca s, which is of no concern here.

Physicists typically call $A_0$ a \emph{phenomenological theory} and refer to $A_{\hbar}$ at $\hbar>0$ 
as a  \emph{fundamental theory}; we sometimes write $A_{>0}$ for the latter family. 
In the philosophy of science, $A_0$ is often referred to as a \emph{higher-level theory} or a   \emph{reduced theory}, in which case  $A_{>0}$   is said to be a
 \emph{lower-level theory},  or a \emph{reducing theory}, respectively. 
 It is important to realize that $A_0$ plays double role: one the one hand, it is
 an idealization or a limiting case of the family of $A_{\hbar}$-theories as $\hbar\raw 0$, whilst on the other 
hand it is defined and understood by itself; indeed, in our examples the $A_0$-theories historically predate the 
$A_{>0}$ theories and were once believed to be real and absolutely correct (if not holy).  Furthermore, in many interesting cases (including the ones above) $A_0$
 \emph{taken by itself} has important features that are surprising or even seem out of the question in its role as a limiting theory. For example, 
 classical mechanics and  thermodynamics both allow spontaneous symmetry breaking, which according to conventional wisdom (corrected below) is absent in any finite quantum system. Moreover, measurements in classical physics  have outcomes, whereas \qm\ faces the infamous measurement problem, see \S\ref{mmp} below.
 
The fact that the phenomenological theory does not really exist in nature, as opposed to the fundamental theory that gives rise to it---or so we assume---poses a severe problem: \emph{Real physical systems are supposed to be described by $A_{>0}$
rather than by the idealization $A_0$, yet in nature these systems display the surprising  features claimed to be intrinsic to $A_0$ (and denied to $A_{>0}$).} For example, the reality of ferromagnetism shows that spontaneous symmetry breaking occurs in nature, although finite materials are supposed to be described by the $A_{>0}$-theory that allegedly forbids it. 
Similarly,  measurements have outcomes, although the world is described by the $A_{>0}$-theory (viz.\ \qm) that fails to predict this. In sum, \emph{the $A_{>0}$-theory fails to describe key features of the real systems that should fall within its scope, whereas the $A_0$-theory  describes these features despite the fact that, being an idealization, real systems in principle do not fall within its scope. }

To resolve this paradox, we introduce two principles that, if adhered to,  should guarantee the  link between theory and reality.
 \emph{Earman's Principle} states that:
\begin{quotation}
`While idealizations are useful and, perhaps, even essential to progress in physics, a sound principle of interpretation would seem to be that no effect  can be counted as  a genuine physical effect if it disappears
when the idealizations are removed.' \hfill (Earman, 2004, p.\ 191)
\end{quotation}
 \emph{Butterfield's Principle} follows up on this, claiming what should happen instead is that:
\begin{quotation}
`There is a weaker,
yet still vivid, novel and robust behaviour  that
occurs before we get to the limit, i.e. for finite $N$. And it is this weaker behaviour
which is physically real.' \hfill (Butterfield, 2011, p.\ 1065)
\end{quotation}
Although these principles should, in our view, be uncontroversial, since the link between theory and reality stands or falls with them, it is remarkable how often idealizations violate them. All rigorous theories of spontaneous symmetry breaking in quantum statistical mechanics strictly apply to infinite systems only (Bratteli \& Robinson, 1981;  Liu \& Emch, 2005), and similarly in quantum field theory (Haag, 1992; Ruetsche, 2011). The same is true for the `Swiss' approach to the measurement problem based on superselection rules (Hepp, 1972;  Emch \& Whitten-Wolfe, 1976), and more recent developments thereof by
 Landsman (1991, 1995) and Sewell (2005). For a change, we here side with Bell (1975)!
    \section{Rethinking the measurement problem}\label{mmp}
 To the best of our knowledge, apart from some very general comments, e.g., in his Como Lecture and in his correspondence (Zinkernagel, 2016), the only time Bohr wrote in some technical detail about  the quantum-mechanical measurement process was in his  papers with Rosenfeld on the measurability of the electromagnetic field (Bohr, 1996, pp.\ 53--166).  However, these comments and papers predate the Cat problem raised by Schr\"{o}dinger (1935),  and do not seem to address what we now see as the \mmp\ of \qm. Indeed, 
 one wonders how Bohr would respond to the amazing recent experiments like Arndt et al (2015), Hornberger et al (2012),  
 Kaltenbaek et al (2015),  Palomaki et al (2013), and  Kovachy et al (2015).
 
 Nonetheless,  Bohrification provides a clear lead on the \mmp. 
 Let us begin with the proper formulation of the problem, on which its  solution should evidently be predicated.
 To set the stage,  Maudlin (1995) distinguishes three \mmp s:
 \begin{enumerate}
\item \emph{The problem of outcomes} states that the following assumptions are contradictory:
\begin{enumerate}
\item The wave-function of the system is complete;
\item The wave-function always evolves linearly (e.g., by the Schr\"{o}dinger equation);
\item Measurements have determinate outcomes.
\end{enumerate}
\item  \emph{The problem of statistics} is that 1(a) = 2(a) and 1(b) = 2(b) also contradict:
\begin{enumerate}
\item[(c)] Measurement situations which are described by identical initial wave-functions sometimes have different outcomes, and the probability of each possible outcome is given by the Born rule.
\end{enumerate}
\item  \emph{The problem of effect} requires that any (physical or philosophical) mechanism producing measurement outcomes should also update the predictions of \qm\ for subsequent measurements (typically that these have the same outcome). 
\end{enumerate}
Furthermore, Maudlin (1995) also gives a classification of potential solutions: 
\begin{itemize}
\item Hidden-variable theories abandon 1(a) = 2(a);
\item Collapse theories abandon 1(b) = 2(b)
\item Multiverse theories abandon 1(c) and reinterpret the word ``different'' in 2(c).
\end{itemize}
As quite rightly emphasized by Maudlin, turning superpositions  into mixtures 
does not even address any of these problems, which is why  decoherence (or, for that matter, superselection rules),
though an interesting prediction of \qm\ (Joos et al, 2003; Schlosshauer, 2007) by itself fails to solve the \mmp, despite occasional but persistent claims to the contrary (especially in the literature on experimental physics).  

Our own take on the \mmp\ is a bit different. First, an apparent problem famously emphasized by Bell (1990) is that ``measurement''  is a \emph{priori} undefined within \qm.  Hence the first part of the measurement problem is to define measurement. In our view, this part of the problem \emph{was} solved by Bohr through his doctrine of classical concepts, which effectively says that a physical interaction \emph{defines} a measurement as soon as the  measurement device---though ultimately quantum-mechanical in an ontological sense---is \emph{described} or \emph{perceived} classically. Thus the doctrine of classical concepts is an epistemological  move (Scheibe, 1973; Camilleri \& Schlosshauer, 2015). 

Without such an interpretation of measurement (which places it outside \qm), there is no \mmp\ in the first place, since there are no outcomes.  
To Bohr, his definition of measurement through the doctrine of classical concepts seems to have been the end of it, but to us, it is just the beginning!  Recalling (from \S\ref{ideal}) that the $A_0$-theory in which measurement outcomes are defined is a limit of the $A_{>0}$-theory responsible for these outcomes, it should be the case that the limits of the underlying (pure) quantum states of the measurement device as $\hbar\raw 0$ (or $N\raw\infty$, either of which limit would enforce a classical description) are pure states on $A_0$, since facts in classical physics correspond to pure (i.e., dispersion-free) states, at least approximately. But this is not at all what \qm\ predicts!  In typical Schr\"{o}dinger Cat situations the limits in question are \emph{mixed} states that are not even approximately pure (Landsman \& Reuvers, 2013; Landsman, 2013). But in reality  sharp outcomes always occur, and so we obtain a violation of  Butterfield's Principle: outcomes (mathematically represented by pure states in $A_0$), are not foreshadowed in $A_{>0}$ (whose superpositions induce mixtures in the limit), although at the same time $A_0$  should arise from $A_{>0}$ as a limit theory. 

With Butterfield's Principle, the link between theory and reality falls. In other words, it is by no means enough that (in Schr\"{o}dinger Cat situations) a classical description of the measurement device turns pure quantum superpositions into mixtures (which is trivially the case in the limit): 
 all terms but one eventually have to \emph{disappear} from the state. According to Butterfield's Principle
this disappearance must (approximately) take place already in \qm\ (if only in the limiting regime); a \emph{Deus ex Machina} that suddenly turns classical mixtures into classical pure states (like the ignorance interpretation of classical probability) but did not previously act on the underlying quantum theory  violates   Earman's Principle, since in reality the classical limit is never reached; cf.\ \S\ref{ideal}. 

To see the mathematics of asymptotic Bohrification in a simple model of  measurement, consider the continuous bundle of \ca s over $I=[0,1]$ that has fibers
 \begin{eqnarray}
A_0&=& C_0(\R^2);\label{A1}\\
A_{\hbar}&=& B_0(L^2(\R))\:\: (\hbar>0), \label{A2}
\end{eqnarray}
and whose continuous cross-sections are defined as follows. First, we recall the well-known \emph{coherent states}
that for  each $(p,q)\in\R^2$ and $\hbar>0$ are given by
\beq
\phi^{(p,q)}_{\hbar}(x)=(\pi\hbar)^{-1/4}e^{-
ipq/2\hbar}e^{ipx/\hbar}e^{-(x-q)^2/2\hbar}.\label{pqcohst} 
\eeq
These states were originally introduced by Schr\"{o}dinger as models for wave packets localized in phase space, and they minimize the Heisenberg uncertainty relations. In terms of these, for each $f\in  C_0(\R^2)$ and $\hbar>0$ we define (compact) operators $\Qh(f)$ on $L^2(\R)$ by
\beq
 Q_{\hbar}(f)=\int_{\R^{2n}} \frac{dp
dq}{2\pi\hbar}\, f(p,q) | \phi^{(p,q)}_{\hbar}\rangle\langle\phi^{(p,q)}_{\hbar}|,
\eeq
where $|\phi\ra\la\phi|$ is the projection onto $\phi$. 
In the literature, $\Qh(f)$ is often called the  \emph{Berezin quantization} of the classical observable $f$.  Finally, the  continuous cross-sections of the bundle with fibers \er{A1} - \er{A2} are 
the maps $\hbar\mapsto\Qh(f)$, with $Q_0(f)=f$; see Landsman (1998) for details.
For our present purposes, the main point is that we may now track down quantum states to classical states, all the way in the limit $\hbar\raw 0$. 
Namely, for each unit vector $\psi\in  L^2(\R)$ the map $f\mapsto \la\ps, Q_{\hbar}(f)\psi\ra$ defines a state on $C_0(\R^2)$, so that by the Riesz--Radon representation theorem we obtain a  probability measure 
 $\mu_{\psi}$ on $\R^2$, given by
\begin{equation}
d\mu_{\ps}(p,q)= \frac{dp
dq}{2\pi\hbar}\, |\la \phi^{(p,q)}_{\hbar},\ps\ra|^2. \label{defmupsi}
\end{equation}
The  probability density $ \chi_{\psi}(p,q)=|\langle \phi^{(p,q)}_{\hbar},\psi\rangle|^2$ is called the \emph{Husumi function} of $\psi$; it gives  a phase space portrait of $\psi$, which is especially useful in studying the limit $\hbar\raw 0$. 

Let us illustrate this formalism for one-dimensional anharmonic oscillator, i.e.,  
\begin{equation}
H_{\hbar}=-\frac{\hbar^2}{2m}\frac{d^2}{dx^2} +\quar \lm (x^2-a^2)^2. \label{TheHam}
\end{equation}
where $\lm>0$, as in the Higgs mechanism. 
It is well  known that the ground state of this Hamiltonian is unique,
despite the fact that the corresponding classical system has two degenerate ground states, given by the phase space points
$(p_0=0,q_0=\pm a)$. In particular, classically this models displays spontaneous symmetry breaking, whereas quantum-mechanically it does not, for any value of $\hbar>0$. This already seems to spell doom for Earman's and Butterfields's Principles (and hence for the link between theory and reality).

A quantitative analysis confirms this threat. The ground state wave-function $\psi_{\hbar}^{(0)}$  is  real and positive definite, and has two peaks, above $x=\pm a$, with exponential decay $|\psi_{\hbar}^{(0)}(x)|\sim \exp( -1/\hbar)$ in the classically forbidden region. In the limit $\hbar\raw 0$ the associated probability measure $\mu_{\hbar}^{(0)}$ converges to the classical mixture
$\half(\dl(p,q-a)+\dl(p,q+a))$:
\begin{center}
\includegraphics[width=0.49\textwidth]{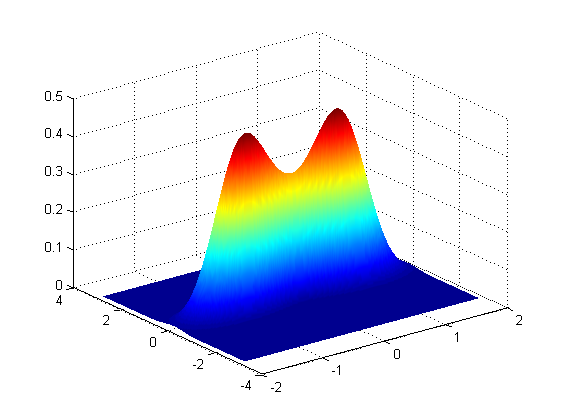}
\includegraphics[width=0.49\textwidth]{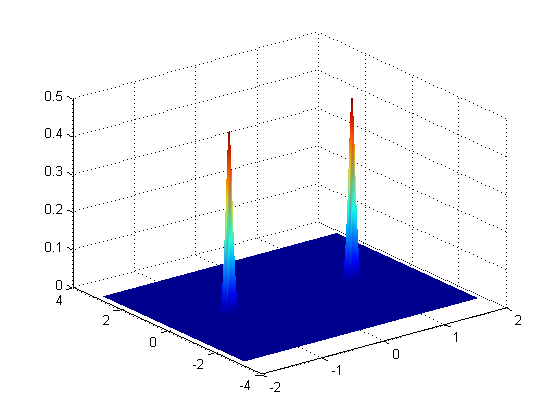}
{Husumi functions of double well ground state for $\psi_{\hbar=0.5}^{(0)}$ (left) and  $\psi_{\hbar=0.01}^{(0)}$ (right).}
\end{center}
 It takes little imagination to see the classical ground states of this systems as pointer states, the (unique) quantum-mechanical ground state
then corresponding to a Schr\"{o}dinger Cat state (Landsman \& Reuvers, 2013). The limit $\hbar\raw 0$, which in the real world (where $\hbar$ is constant) corresponds to $m\raw\infty$ or $\lm\raw\infty$, or even to $N\raw\infty$, as in the closely related but superior model of Spehner \& Haake (2008),  fails to predict the empirical fact that in reality the system (typically a particle) is found at one of the two minima (i.e., the cat is either dead or alive), and hence  Earman's and Butterfields's Principles are indeed violated:
the quantum theory ought to have foreshadowed the symmetry breakdown of the limiting classical theory, but it fails to do so however close it is to the classical limit. 

Motivated by examples like this---cf.\ Landsman (2013) for some others---we are led to a  formulation of the \mmp\ that is a bit different from the ones above:
\begin{enumerate}
\item[4.] \emph{Problem of classical outcomes.} The following assumptions are contradictory:
\begin{enumerate}
\item Outcomes of measurements on quantum systems are classical;
\item Classical physics is an appropriate limit of quantum physics.
\item Quantum states are dynamically unaffected by this limit.
\end{enumerate}
\end{enumerate}

Compared with Maudlin's three \mmp s, the most striking difference with the formulation just given is our mere implicit reference to the linear nature of (unitary) time-evolution in \qm, through the word ``dynamical'' in (c). Moreover, all reference to the possible completeness of quantum states has also, implicitly, been bracketed under (c). The reason  is that Maudlin's (and everyone else's) assumption 1(b), which  forms the basis of all discussions of the \mmp, is actually a counterfactual of the kind one should avoid in \qm\ (van Heugten \& Wolters, 2016). What this assumption says is that if $\psi_D$ \emph{were} the initial state of Schr\"{o}dinger's Cat, then it \emph{would} evolve (linearly) according to the  Schr\"{o}dinger equation with \emph{given} Hamiltonian $h$. If the initial state  \emph{were} $\psi_L$, then it  \emph{would} evolve according to the \emph{same} Hamiltonian $h$, and 
if it  \emph{were} $(\ps_D+\ps_L)/\sqrt{2}$, then it  \emph{would} evolve according to $h$, too.

 However, a fundamental lesson from the models in Landsman \& Reuvers (2013)  is that in typical measurement situations
unitary quantum dynamics is extremely unstable in the classical limit $\hbar\raw 0$. See Jona-Lasinio, Martinelli, \& Scoppola (1981) and Simon (1985) for the original mathematical arguments to this effect (in our view these two papers belong to the most important ever written about the foundations of \qm, although they were not intended as such), as well as
 Landsman (2013) for similar results in the limit $N\raw\infty$. 
 This implies that at least in the classical regime of \qm\ the counterfactual linearity assumption underlying the usual discussions of the \mmp\  is meaningless, because very slight modifications to the Hamiltonian (which are harmless in the quantum regime), whilst keeping the dynamics unitary, suffice to destabilize the wave-function. For example, as $\hbar\raw 0$ a tiny asymmetric perturbation to the double well potential is enough to localize the quantum ground state, and this happens in accordance with Butterfield's Principle, as the effect already occurs for small $\hbar>0$. Similarly, ground states of spin chains that are of  Schr\"{o}dinger Cat type for small $N$ localize (in spin configuration space) under tiny 
  asymmetric perturbation of the Hamiltonian as $N\raw\infty$, and this already happens for large but finite $N$ (Landsman, 2013). 
 
 Thus we envisage  the solution to the \mmp\ (at least in the above reformulation) to be as follows:
 the task being to find a dynamical mechanism that removes all but one terms of a typical post-measurement superposition of quantum states, the key \emph{conceptual} point is that according to Bohr this task only needs to be accomplished near the classical limit, upon which they key \emph{technical} point is that asymmetric perturbations to the Hamiltonian, whilst having a negligible effect in the quantum regime, destabilize the superposition so as to leave just one term in the classical regime. Thus the dynamical mechanism is effective where it should, and does not operate where it should not. 
In response to Leggett  (2002), the eventual collapse of macroscopic Schr\"{o}dinger Cat states does not require the
demise of quantum theory, as this collapse is guaranteed by the instability under tiny perturbations of quantum dynamics in the classical regime. 

In conclusion, of our formulation no.\ 4 of the \mmp\ we reject assumption (c), effectively proposing a (dynamical) collapse theory of a new kind (both technically and conceptually). 
 This also explains 2(c) and solves 3. Assumption 1(a) is somewhat irrelevant, since what is incomplete in practice is not the state but the Hamiltonian (since its tiny perturbations are typically hidden). As explained above,  Assumption 1(b) is a dangerous counterfactual, so we do not make it, whereas we wholeheartedly endorse 1(c)! 
 
 Having said this, many problems remain to be overcome since the first steps were taken in Landsman \& Reuvers (2013)
 and Landsman (2013); see van Heugten \& Wolters (2016).
\section*{Appendix: The mathematics of asymptotic Bohrification}\addcontentsline{toc}{section}{Appendix: The mathematics of asymptotic Bohrification} 
Asymptotic Bohrification is based on the notion of a \emph{continuous bundle of \ca s}. This concept may sound unnecessarily technical, but it provides the  mathematical bridge between the classical and the quantum and hence deserves to be more widely known.

 Let $I$ be a compact space, which for us is always a
subset of the unit interval $[0,1]$ that  contains 0 as an accumulation point, so one may have e.g.\ $I=[0,1]$ itself, or $I=(1/\mathbb{N})\cup \{0\}\equiv1/\dot{\N}$, where $\mathbb{N}=\{1,2,\ldots\}$. In asymptotic Bohrification, $I$ either plays the role of the value set for Planck's constant, in which case one is interested in the limit $\hbar\raw 0$, 
or it indexes $1/N$, where $N\raw\infty$ is the number of particles in some system, or the number of experiments (as in our derivation of the Born rule in \S\ref{Born}), or the principal quantum number in atomic orbit theory (as in Bohr's original correspondence principle). Also in the latter case we generically write $\hbar\in I$, so that $N=1/\hbar$. 
A \emph{continuous bundle of $C^*$-algebras} over $I$, then,  consists of a
 \ca\ $A$, a collection of \ca s $(A_{\hbar})_{{\hbar}\in I}$, and surjective homomorphisms $\phv_{\hbar}:A\raw A_{\hbar}$ for each $\hbar\in I$ , such that:
\begin{enumerate}
\item
The function ${\hbar}\mapsto  \|\phv_{\hbar}(a)\|_{\hbar}$ is in $C(I)$ for each $a\in A$ (in particular at $\hbar=0$!).
\item
Writing  $\|\cdot\|_{\hbar}$ for the norm in $A_{\hbar}$, 
the norm of any $a\in A$ is given by
\begin{equation}
\| a\|=\sup_{{\hbar}\in I}\|\phv_{\hbar}(a)\|_{\hbar}.
\end{equation}
\item
For any $f\in C(I)$ and $a\in A$ there is an element $fa\in A$ such that for each ${\hbar}\in I$,  
\beq
\phv_{\hbar}(fa)=f({\hbar})\phv_{\hbar}(a).
\eeq
\end{enumerate}

 A  \emph{continuous (cross-) section} of such a bundle is a map $\hbar\mapsto a(\hbar)\in A_{\hbar}$, $\hbar\in I$,  for which there is an $a\in A$ such that $a(\hbar)=\phv_{\hbar}(a)$ for each ${\hbar}\in I$. 
Thus the \ca\ $A$ may  be identified with the space of continuous sections of the bundle: if we do so, the homomorphism $\phv_{\hbar}$ is just the evaluation map at $\hbar$.  The structure of $A$  as a \ca\ then corresponds to pointwise operations on sections. 
The idea is that the  family $(A_{\hbar})_{\hbar\in I}$ of \ca s is glued together by specifying a topology on the
disjoint union $\sqcup_{\hbar\in [0,1]}A_{\hbar}$, seen as a fibre
 bundle over $I$.
 However, this topology is in fact given rather indirectly, namely via the specification of the space of continuous sections.

Another way to look at continuous bundles of \ca s is to start from a non-degenerate homomorphism $\phv$ from $C(I)$ to the center $Z(M(A))$ of the multiplier algebra $M(A)$ of $A$; we  write  $fa$ for $\phv(f)a$ (in this notation non-degeneracy means that $C(I)A$ is dense in $A$, which implies
 that $C(I)A=A$).
 Given $\phv$,  one may define  fiber algebras by 
 \begin{eqnarray}
 A_{\hbar}&=&A/(C_0(I;\hbar)\cdot A);\\
 C_0(I;\hbar)&=&\{f\in C(I)\mid f(\hbar)=0\},
\end{eqnarray}
 and $C_0(I;\hbar)\cdot A$ is an ideal in $A$, making the quotient $ A_{\hbar}$ a \ca. 
The projections $\phv_{\hbar}:A\raw A_{\hbar}$ are then given by the corresponding quotient maps (sending $a\in A$ to its equivalence class in $A_{\hbar}$). Since in general the function 
  ${\hbar}\mapsto  \|\phv_{\hbar}(a)\|_{\hbar}$  is merely upper semicontinuous, one only obtains a structure  equivalent to the one described in the above definition if one explicitly requires the above function to be in $C(I)$, in which case 
  property 2 follows, too. See e.g.\ Williams (2007, App.\ C) for the material in this appendix. 
\section*{Acknowledgement}  
Research on Bohrification has been  supported by the Radboud University, the Netherlands Organization for Scientific Research (NWO), and the Templeton World Charity Foundation (TWCF).
The author was also very fortunate in having been surrounded by such good (PhD) students  and postdocs contributing to the Bohrification program: in alphabetical order these were
Martijn Caspers, Ronnie Hermens, Jasper van Heugten, Chris Heunen, Bert Lindenhovius, Robin Reuvers, Bas Spitters, Marco Stevens, and Sander Wolters. 
\section*{References} \addcontentsline{toc}{section}{References} 
\begin{footnotesize}
\begin{trivlist}
\item  Aarens, J.F. (1970).  Quasi-states on \ca s. \emph{Transactions of the American Mathematical Society} 149, 601--625.
 \item Anderson, J. (1979). Extensions, restrictions, and representations of states on C*-algebras.
\emph{Transactions of the American Mathematical Society} 249, 303--329. 
\item  Arndt, M., D\"{o}rre, S.,  Eibenberger, S., Haslinger, P.,  Rodewald, J.,  Hornberger, K.,  Nimmrichter, S.,  Mayor, M. (2015).
Matter-wave interferometry with composite quantum objects. \texttt{arXiv:1501.07770}.
\item  Banaschewski, B., Mulvey,  C.J. (2006). 
A globalisation of the {G}elfand duality theorem. \emph{ Annals of Pure and Applied Logic} 137, 62--103. 
\item Bayen, F., Flato, M., Fronsdal, C.,  Lichnerowicz, A., 
Sternheimer, D.  (1978).  Deformation theory and quantization I, II.  \emph{Annals of  Physic (N.Y.)} 110, 61--110, 111--151.
 \item 
Bell, J.S. (1966).   On the Problem of hidden variables in quantum mechanics.
 \emph{Reviews of Modern Physics} 38, 447--452.
  \item 
Bell, J.S. (1975). 
On wave packet reduction in the Coleman-Hepp model. \emph{Helvetica Physica Acta} 48, 93--98.
\item Bell, J.S. (1990). Against ``Measurement''.  \emph{Sixty-Two Years of Uncertainty}, Ed. Miller, A.I., pp.\ 17--31.
 \item Belinfante, F.J. (1973). \emph{A Survey of Hidden-Variable Theories}. Oxford: Pergamon Press.
 \item
Berezin, F.A. (1975). General concept of quantization.  \emph{Communications in
Mathematical Physics}  40, 153--174.
\item
 Birkhoff,  G., Neumann,   J. von (1936).
The logic of quantum mechanics. \emph{Annals of Mathematics} {37},  823--843.
\item Bohr, N. (1949). Discussion with Einstein on epistemological problems in atomic physics. \emph{Albert Einstein: Philosopher-Scientist},  Ed. Schlipp, P.A.,  pp.\ 201--241.  La Salle: Open Court. 
\item Bohr, N. (1976).   \emph{Collected Works. Vol.\ 3:
The Correspondence Principle (1918--1923).} Eds. Rosenfeld, J., Nielsen, R.  Amsterdam: North-Holland.
\item Bohr, N. (1985). \emph{Collected Works. Vol.\ 6: Foundations of Quantum Physics {\sc i} (1926--1932)}. Ed. Kalckar, J. Amsterdam: North-Holland.
\item Bohr, N. (1996). \emph{Collected Works. Vol.\ 7: Foundations of Quantum Physics {\sc ii} (1933--1958)}. Ed. Kalckar, J.  Amsterdam: North-Holland.
\item Bokulich, A.(2008). \emph{Reexamining the Quantum-Classical Relation: Beyond Reductionism and Pluralism}.
Cambridge: Cambridge University Press. 
\item  Bratteli, O., Robinson, D.W. (1981). \emph{Operator Algebras and Quantum Statistical Mechanics. Vol.\ II:
Equilibrium States, Models in Statistical Mechanics}. Berlin: Springer.
\item  
 Bub, J. (2011). Is von Neumann's `no hidden variables' proof silly?,
 \emph{Deep Beauty: Mathematical Innovation and the Search for Underlying Intelligibility in the Quantum World}, Ed.
  Halvorson, H.,  pp.\ 393--408. Cambridge: Cambridge University Press. 
  \item Butterfield, J. (2011). Less is different: Emergence and reduction reconciled.
\emph{Foundations of Physics} 41, 1065--1135.
\item  Camilleri, K. (2007). Bohr, Heisenberg and the divergent views of complementarity. 
\emph{Studies in History and Philosophy of Modern Physics} 38, 514--528.
%\item Camilleri, K. (2009). \emph{Heisenberg and the Interpretation of Quantum Mechanics: The Physicist as Philosopher}. %Cambridge: Cambridge University Press. 
\item Camilleri, K., Schlosshauer, M. (2015). Niels Bohr as philosopher of experiment: Does decoherence theory challenge Bohr's doctrine of classical concepts?
\emph{Studies in History and Philosophy of Modern Physics} 49, 73--83. 
\item Caruana, L. (1995). John von Neumann's `Impossibility Proof' in a historical perspective.
\emph{Physis} 32, 109--124.
%\item  Casazza, P.G.,  Fickus, M.,  Tremain, J.C.,  Weber, E. (2005).
% The Kadison-Singer Problem in mathematics and engineering.
 %\texttt{arXiv:math/0510024}.
\item  Caspers, M., Heunen, C., Landsman, N.P., Spitters, B. (2009). 
Intuitionistic quantum logic of an $n$-level system. \emph{Foundations of Physics} 39, 731--759.
\item
 Cassinello, A., S\'{a}nchez-G\'{o}mez, J.L. (1996).
 On the probabilistic postulate of quantum mechanics. \emph{Foundations of  Physics} 26, 1357--1374. 
\item Caves, C.,  Schack,  R. (2005). Properties of the frequency operator do not imply the quantum probability postulate.
 \emph{Annals of Physics (N.Y.)}
 315, 123--146.
\item Connes, A. (1994). \emph{Noncommutative
Geometry}. San Diego: Academic Press. 
\item Darrigol, O. (1992). \emph{From c-Numbers to q-Numbers}.  Berkeley: University of California Press. 
\item Dirac, P.A.M. 1930). \emph{The Principles of Quantum
Mechanics}. Oxford: Clarendon Press. 
\item  D\"{o}ring, A. (2005). Kochen--Specker Theorem for von Neumann algebras.
\emph{International Journal of Theoretical Physics} 44, 139--160.
\item  D\"{o}ring, A. (2012). Topos-based logic for quantum systems and Bi-Heyting algebras.
\emph{Logic \& Algebra in Quantum Computing, Lecture Notes in Logic}, in press. Cambridge: Cambridge University Press.
\texttt{arXiv:1202.2750}.
\item D\"{o}ring, A., (2014). Two new complete invariants of von Neumann algebras.
\texttt{arXiv:1411.5558}.
\item D\"{o}ring, A., Harding, J. (2010). Abelian subalgebras and the Jordan structure of a von Neumann algebra.
\texttt{arXiv:1009.4945}.
\item D\"{o}ring, A., Isham, C.J. (2010). ``What is a Thing?'': Topos theory in the foundations of physics. 
\emph{Lecture Notes in Physics} 813, 753--937.
\item Earman, J. (2004). Curie's Principle and spontaneous symmetry breaking. \emph{International Studies in the Philosophy of Science} 18, 173--198. 
%\item Eibenberger, S.,  Gerlich, S., Arndt, M., Mayor, M., T\"{u}xen, J. (2013). Matter-wave interference of particles selected from a molecular library with %masses exceeding 10 000 amu. \emph{
%Physical Chemistry Chemical Physics} 35, 14696--14700.
\item Eilers, M., Horst, E. (1975). The theorem of Gleason for nonseparable Hilbert spaces.
\emph{International Journal of Theoretical Physics} 13, 419--424.
\item  Emch, G.G.,  Whitten-Wolfe, B. (1976). Mechanical quantum measuring process.
 \emph{Helvetica Physica Acta}  49, 45--55.
\item 
 Farhi,  E.,  Goldstone, J.,  Gutmann, S. (1989). How probability arises in quantum mechanics.
 \emph{Annals of Physics (N.Y.)} 192, 368--382. 
\item   Faye, J. (2002). Copenhagen Interpretation of Quantum Mechanics.
\emph{The Stanford Encyclopedia of Philosophy (Summer 2002 Edition)}. Ed.  Zalta, E.N.Ê \\ \texttt{http://plato.stanford.edu/archives/sum2002/entries/qm-copenhagen/}.    
\item
 Fine, A. (1974). On the completeness of quantum theory. \emph{Synthese} 29, 257--289.
 \item  Finkelstein, D. (1965). The logic of quantum physics. \emph{Transactions of the New York Academy of Science} 25, 621--637.  
\item Folse, H.J. (1985). \emph{ The Philosophy of Niels Bohr}. Amsterdam: North-Holland.
\item  Gelfand, I.M., Naimark, M.A. (1943). On the imbedding of normed rings into the ring of operators in Hilbert space.
\emph{Sbornik: Mathematics} 12, 197--213.
\item Gleason, A.M. (1957). Measures on the closed subspaces of a \Hs. \emph{Journal of Mathematics and Mechanics}
6, 885--893.
\item Gottwald, S. (2015).
 Many-Valued Logic. \emph{The Stanford Encyclopedia of Philosophy (Spring 2015 Edition)}, Ed. Zalta, E.N. 
 \verb#http://plato.stanford.edu/archives/spr2015/entries/logic-manyvalued/#.
\item   Haag, R. (1992).   \emph{Local Quantum Physics: Fields, Particles, Algebras}. 
Heidelberg: Springer-Verlag.  
\item  Hamhalter, J. (1993).  Pure Jauch-Piron states on von Neumann algebras.
\emph{Annales de l'IHP Physique th\'{e}orique} 58, 173--187.
\item Hamhalter, J. (2004). \emph{Quantum Measure Theory}. Dordrecht: Kluwer Academic Publishers.
\item  Hamhalter,  J. (2011).
Isomorphisms of ordered structures of abelian C*-subalgebras of C*-algebras,
\emph{Journal of Mathematical Analysis and Applications} 383, 391--399.
\item  Hamhalter,  J.  Turilova, E. (2013). Structure of associative subalgebras of Jordan operator algebras.
\emph{Quarterly Journal of Mathematics}  64,  397--408.
\item Hamilton, J., Isham, C.J.,  Butterfield, J. (2000).
Topos perspective on the Kochen--Specker Theorem: III. Von Neumann Algebras as the base category.
 \emph{International Journal of Theoretical Physics} 39, 1413--1436.
 \item
Hartle, J.B. (1968). Quantum mechanics of individual systems. \emph{American Journal of Physics} 36, 704--712.  
\item Heisenberg, W. (1958). \emph{Physics and Philosophy: The Revolution in Modern Science}. London: Allen \&\ Unwin.
 \item Hendry, J. (1984). \emph{The Creation of Quantum Mechanics and the Bohr-Pauli Dialogue}.  Dordrecht: Reidel.
  \item  Hepp, K.  (1972). Quantum theory of measurement and macroscopic observables.
  \emph{Helvetica Physica Acta}  45, 237--248.
\item Hermann, G. (1935). Die naturphilosphischen Grundlagen der Quantenmechanik.
 \emph{Abhandlungen der Fries'schen Schule} 6, 75--152. 
 \item 
Hermens, R. (2009). \emph{Quantum Mechanics: From Realism to Intuitionism}. M.Sc Thesis, Radboud University Nijmegen.
\texttt{http://philsci-archive.pitt.edu/5021/}.
 \item 
Hermens, R. (2016). \emph{Philosophy of Quantum Probability}. PhD Thesis, Rijksuniversiteit Groningen. 
 \item Heugten, J. van, Wolters, S. (2016). Obituary for a flea.
  \emph{Proceedings of the Nagoya Winter Workshop 2015: Reality and Measurement in Algebraic Quantum Theory}, Ed. Ozawa, M. In preparation.
\item  Heunen, C. (2009). \emph{Categorical Quantum Models and Logics}.   PhD Thesis, Radboud University Nijmegen.
\item  Heunen, C.  (2014a). Characterizations of categories of commutative C*-algebras. \emph{Communications in Mathematical
Physics} 331, 215--238. 
\item  Heunen, C.  (2014b). The many classical faces of quantum structures. \texttt{arXiv:1412.2177}.
\item  Heunen, C., Landsman, N.P., Spitters, B. (2009). 
A topos for algebraic quantum theory. \emph{Communications in Mathematical Physics} 291, 63--110.
\item  Heunen, C., Landsman, N.P., Spitters, B. (2012).
 Bohrification of operator algebras and quantum logic.  \emph{Synthese}, 186, 719--752.
  \item  Heunen, C., Landsman, N.P., Spitters, B., Wolters, S. (2012).
  The Gelfand spectrum of a noncommutative C*-algebra: a topos-theoretic approach.
  \emph{Journal of the Australian Mathematical Society} 90, 32--59.
    \item  Heunen, C., Lindenhovius, A.J. (2015). Domains of commutative C*-subalgebras.
   \texttt{arXiv:1504.02730}.   
\item Hornberger, K., Gerlich, S.,  Haslinger, P.,  Nimmrichter,S.,  Arndt, M. (2012).
Colloquium: Quantum interference of clusters and molecules. \emph{Reviews of Modern Physics} 84, 157--173.
\item Howard, D. (2004). Who Invented the Copenhagen Interpretation? 
\emph{Philosophy of Science} 71, 669--682.
\item Isham, C.J.,  Butterfield, J. (1998). Topos perspective on the Kochen--Specker theorem: I. Quantum states as generalized valuations. \emph{International Journal of Theoretical Physics} 37, 2669--2733. 
\item Ivrii, V. (1998). \emph{Microlocal Analysis and Precise Spectral Asymptotics}. New York: Springer-Verlag.
\item
Jona-Lasinio, G., Martinelli, F., Scoppola, E. (1981).
 New approach to the semiclassical limit of quantum mechanics.
\emph{Communications in Mathematical Physics} 80, 223--254.
\item Joos, E., Zeh, H.D., Kiefer, C., Giulini, D., Kupsch, J.,  Stamatescu, I.-O. (2003).
\emph{Decoherence and the Appearance of a Classical World in Quantum Theory. Second Edition}. Berlin: Springer-Verlag.
\item Kadison, R.V.,  Ringrose, J.R. (1983). \emph{Fundamentals of the Theory of Operator Algebras. Vol. 1: Elementary Theory}.   New York: Academic Press.
 \item Kadison, R.V., Ringrose, J.R. (1986). \emph{Fundamentals of the Theory of Operator Algebras. Vol. 2: Advanced  Theory}.   New York: Academic Press. 
\item Kadison, R.V.,  Singer, I.M. (1959). Extensions of pure states. \emph{American Journal of Mathematics} 81, 383--400. 
%\item Kalmbach, G. (1998). \emph{Quantum Measures and Spaces}. Dordrecht: Springer. 
\item Kaltenbaek, R. et al (2015). Macroscopic quantum resonators (MAQRO): 2015 Update.  \texttt{arXiv:1503.02640}. 
 \item Kochen, S.,  Specker, E. (1967).
The problem of hidden variables in quantum mechanics.  \emph{Journal of Mathematics and Mechanics} 17, 59--87.
\item Kovachy, T.,  Asenbaum,	P., Overstreet,  C., Donnelly,  C.A.,  Dickerson,	S.M., 
Sugarbaker, A.,  Hogan, J.M.,  Kasevich, M.A. (2015). Quantum superposition at the half-metre scale.
\emph{Nature} 528, 530--533. 
\item Landsman, N.P. (1991). Algebraic
theory of superselection sectors and the measurement problem in
quantum mechanics. \emph{International Journal of Modern Physics}
A6, 5349--5372. 
\item Landsman, N.P. (1995).  Observation and superselection in quantum mechanics.
\emph{Studies in History and Philosophy of Modern Physics}  26, 45--73.
\item Landsman, N.P. (1998). \emph{Mathematical Topics Between Classical and Quantum Mechanics}. New York: Springer-Verlag. 
\item Landsman, N.P. (2007).
 Between classical and quantum. , \emph{Handbook of
the Philosophy of Science. Vol. 2: Philosophy of Physics, Part A}, Butterfield, J., Earman, J., pp. 417--553.
Amsterdam: North-Holland. 
 \item Landsman, N.P. (2008).
  Macroscopic observables and the Born rule. \emph{Reviews in Mathematical Physics} 20, 1173--1190.
 \item Landsman, N.P. (2009). The Born rule and its interpretation.
  \emph{Compendium of Quantum Physics}, Eds. Greenberger, D., Hentschel, K., Weinert, F., pp. 64--70. Dordrecht: Springer.
 \item   Landsman, N.P., Reuvers, R. (2013).  A flea on Schr\"{o}dinger's Cat. \emph{Foundations of Physics} 43, 373--407.
 \item   Landsman, N.P. (2013).  Spontaneous symmetry breaking in quantum systems: Emergence or reduction?
\emph{Studies in History and Philosophy of Modern Physics} 44, 379--394.  
\item  Landsman, N.P., Lindenhovius, A.J. (2016). Exact Bohrification. \emph{Proceedings of the Nagoya Winter Workshop 2015: Reality and Measurement in Algebraic Quantum Theory}, Ed. Ozawa, M.  In preparation.
\item Landau, L.D.,  Lifshitz, E.M. (1977). \emph{Quantum Mechanics: Non-relativistic Theory}.
3d Ed. Oxford: Pergamon Press.
\item Leggett, A.J. 2002). Testing the limits of quantum mechanics: motivation, state of play,
prospects. \emph{Journal of Physics: Condensed Matter} 14, R415--R451. 
\item Lindenhovius, B. (2015). Classifying finite-dimensional C*-algebras by posets of their commutative C*-subalgebras.
\texttt{arXiv:1501.03030}.
\item Lindenhovius, B. (2016). $\mathcal{C}(A)$.  PhD Thesis, Radboud University Nijmegen.
\item  Liu, C., Emch, G.G.  (2005). Explaining quantum spontaneous symmetry breaking.
\emph{Studies in History and Philosophy of Modern Physics}  36, 137--163. 
\item Mac Lane, S.,   Moerdijk, I. (1992). \emph{Sheaves in Geometry and Logic: A First Introduction to Topos Theory}.
New York: Springer. 
\item Mackey, G.W. (1957). Quantum mechanics and Hilbert space. \emph{ American Mathematical Monthly}
64, 45--57.
\item Maeda, S. (1990). Probability measures on projections in von Neumann algebras.
      \emph{Reviews in Mathematical Physics} 1, 235--290.
 \item  Marcus, A.,  Spielman, D.A.,  Srivastava, N. (2014a). Interlacing families II: Mixed characteristic polynomials and the Kadison--Singer Problem.  \texttt{arXiv:1306.3969}.
  \item  Marcus, A.,  Spielman, D.A.,  Srivastava, N. (2014b).
 Ramanujan Graphs and the solution of the Kadison--Singer Problem.  \texttt{arXiv:1408.4421}.
 \item Martinez, A. (2002). \emph{An Introduction to Semiclassical and Microlocal Analysis}. New York: Springer-Verlag. 
 \item Maudlin, T. (1995). Three measurement problems. \emph{Topoi} 14, 7--15. 
\item  Mehra, J.,  Rechenberg, H. (1982). \emph{The Historical Development of Quantum Theory. Vol.\ 1: The Quantum Theory of Planck, Einstein, Bohr, and Sommerfeld: Its Foundation and the Rise of Its Difficulties}. New York: Springer-Verlag.
\item Moschovakis, J. (2015). Intuitionistic Logic. \emph{The Stanford Encyclopedia of Philosophy (Spring 2015 Edition)}, 
Ed.  Zalta, E.N.  \verb#http://plato.stanford.edu/archives/spr2015/entries/logic-intuitionistic/#.
\item Murray, F.J., Neumann, J. von (1936).  On rings of operators. \emph{Annals of Mathematics}
37, 116--229.
\item Murray, F.J., Neumann, J. von (1937).  On rings of operators {\sc ii}.  \emph{Transactions of the American Mathematical Society} 41, 208--248. 
\item Murray, F.J., Neumann, J. von (1943).  On rings of operators {\sc iv}.
 \emph{Annals of Mathematics} 44, 716--808.
  \item Neumann, J. von (1929). Zur Algebra der Funktionaloperatoren und der Theorie der normalen Operatoren.
  \emph{Mathematische Annalen} 102, 370--427. 
 \item Neumann, J. von (1931). \"{U}ber Funktionen von Funktionaloperatoren. \emph{Annals of Mathematics} 32, 191--226.
\item Neumann, J. von (1932). \emph{Mathematische Grundlagen der Quantenmechanik.}
Berlin: Springer--Verlag. Transl.\ \emph{Mathematical Foundations of Quantum Mechanics}. Princeton: Princeton University Press  (1955).
\item Neumann, J. von (1938). On infinite direct products. \emph{Compositio Mathematica} 6, 1--77. 
\item Neumann, J. von (1940). On rings of operators, {\sc iii}. \emph{Annals of Mathematics} 41, 94--161.
\item Neumann, J. von (1949). On rings of operators, {\sc v}.  Reduction theory.   \emph{Annals of Mathematics} 50, 401--485.
\item Norton, J.D. (2012). Approximation and idealization: Why the difference matters.
\emph{Philosophy of Science} 79, 207--232.
%  \item   Omn\`{e}s, R. (1994). \emph{The Interpretation of Quantum Mechanics}. 
%Princeton: Princeton University Press.
\item Palomaki, T.A.,  Teufel, J.D.,  Simmonds,  R.W.,  Lehnert, K.W. (2013). Entangling mechanical motion with microwave fields. \emph{Science}  342 , no. 6159, 710--713.  
\item Pedersen, G.K. (1989). \emph{Analysis Now}, 2nd ed. New York: Springer-Verlag. 
\item Rieffel, M.A. (1989). Deformation quantization of Heisenberg manifolds.  \emph{Communications in Mathematical Physics}  122, 531--562. 
\item Rieffel, M.A. (1994). 
  Quantization and $C\sp *$-algebras.   \emph{Contemporary Mathematics} 167,  66--97.
   \item Ruetsche, L. (2011). \emph{Interpreting Quantum Theories.} Oxford: Oxford University Press.
  \item Scheibe, E. (1973). \emph{The Logical Analysis of Quantum Mechanics}. Oxford: Pergamon Press.
\item Schlosshauer, M. (2007). \emph{Decoherence and the Quantum-to-Classical Transition}. Berlin: Springer. 
\item  Schr\"{o}dinger, E. (1935). Die gegenw\"{a}rtige Situation in der Quantenmechanik. \emph{Die Naturwissenschaften} 23, 807--812, 823--828, 844--849.
\item Seevinck, M.P. (2012). Challenging the gospel:
Grete Hermann on von Neumann's no-hidden-variables proof. Slides available at
\verb#http://mpseevinck.ruhosting.nl/seevinck/Aberdeen_Grete_Hermann2.pdf#.
\item   Sewell, G.L. (2005).  On the mathematical structure of quantum measurement theory.
\texttt{arXiv:math-ph/0505032v2}.
\item   Simon, B. (1985).
Semiclassical analysis of low lying eigenvalues. IV. The flea on the elephant.  \emph{Journal of Functional Analysis}
 63, 123--136.
\item  Spehner, D., Haake,  F. (2008). Quantum measurements without macroscopic superpositions.
\emph{Physical Review A}  77, 052114.
 \item Stevens, M. (2015). \emph{The Kadison--Singer Property}. MSc Thesis, Radboud University Nijmegen. 
 \item Tao, T. (2013). Real stable polynomials and the Kadison-Singer problem. 
\texttt{http://terrytao.wordpress.com/}\\ \texttt{2013/11/04/real-stable-polynomials-and-the-kadison-singer-problem/}.
\item  Van Wesep, R.A. (2006).  Many worlds and the appearance of probability in quantum mechanics.
 \emph{Annals of Physics (N.Y.)} 321, 2438--2452. 
\item Wallace, D. (2012). \emph{The Emergent Multiverse: Quantum Theory According to the Everett Interpretation.}
Oxford: Oxford University Press. 
\item  Weaver, N. (2004). The Kadison-Singer problem in discrepancy theory. \emph{Discrete Mathematics} 278, 227--239.
\item Williams, D.P. (2007). \emph{Crossed Products of C*-Algebras}. Providence: American Mathematical Society. 
\item  Wolters, S. (2013). \emph{Quantum Toposophy}. PhD Thesis, Radboud University Nijmegen. 
\item Zinkernagel, H. (2016). Niels Bohr on the wave function and the classical/quantum divide.
\emph{Studies in History and Philosophy of Modern Physics}  53,  9--19.
\end{trivlist}
\end{footnotesize}
\end{document}

%% file: Bohrificationpreamble.tex
\newcommand{\quar}{\mbox{\footnotesize $\frac{1}{4}$}}
\newcommand{\half}{\mbox{\footnotesize $\frac{1}{2}$}}

\newcommand{\qm}{quantum mechanics}
\newcommand{\er}{\eqref}
 
\newcommand{\beq}{\begin{equation}}
\newcommand{\eeq}{\end{equation}} 
\newcommand{\bea}{\begin{eqnarray}}
\newcommand{\eea}{\end{eqnarray}}

\newcommand{\ovl}{\overline}
\newcommand{\ul}{\underline}
\newcommand{\wed}{\wedge}
 
\newcommand{\raw}{\rightarrow}

\newcommand{\ot}{\otimes} 
\newcommand{\la}{\langle} \newcommand{\ra}{\rangle}

\newcommand{\x}{\times}

\newcommand{\ca}{C*-algebra}

\newcommand{\Hs}{Hilbert space}

\newcommand{\hv}{hidden variable}

   \newcommand{\vna}{von
Neumann algebra}

\newcommand{\CA}{\mathcal{C}(A)} 
\newcommand{\dl}{\delta} 
 \newcommand{\varep}{\varepsilon}

 \newcommand{\kp}{\kappa}
\newcommand{\lm}{\lambda} 
\newcommand{\rh}{\rho} \newcommand{\sg}{\sigma}
\newcommand{\Sg}{\Sigma}  
 \newcommand{\phv}{\varphi}
\newcommand{\ch}{\ch} \newcommand{\ps}{\psi} 
\newcommand{\om}{\omega} 
\newcommand{\ups}{\upsilon}
\newcommand{\Qh}{Q_{\hbar}}
\newcommand{\inv}{^{-1}}
\newcommand{\Tr}{\mbox{\rm Tr}\,}

\newcommand{\GS}{\mathfrak{S}}

\newcommand{\CO}{{\mathcal O}} \newcommand{\CP}{{\mathcal P}}

\newcommand{\C}{{\mathbb C}} 
\newcommand{\N}{{\mathbb N}} \newcommand{\R}{{\mathbb R}}
 
  \makeatletter
 
\makeatletter
\def\moverlay{\mathpalette\mov@rlay}
\def\mov@rlay#1#2{\leavevmode\vtop{%
   \baselineskip\z@skip \lineskiplimit-\maxdimen
   \ialign{\hfil$\m@th#1##$\hfil\cr#2\crcr}}}
\newcommand{\charfusion}[3][\mathord]{
    #1{\ifx#1\mathop\vphantom{#2}\fi
        \mathpalette\mov@rlay{#2\cr#3}
      }
    \ifx#1\mathop\expandafter\displaylimits\fi}
\makeatother